\documentclass[apj,iop]{emulateapj}
\usepackage{natbib}
\usepackage{graphicx}

\usepackage{tikz}
\usepackage{bm}
\usepackage{amsmath}
\usepackage{amsfonts}
\usepackage{enumitem}
\usepackage{float}
\usetikzlibrary{shapes,arrows}

\tikzstyle{obsable} = [draw, ellipse, text width=2.8em, 
    text centered, minimum height=2.4em, fill=black!30]
\tikzstyle{line} = [draw,  very thick, color=black!50, -latex']
\tikzstyle{param} = [draw, ellipse, text centered, text width=2.8em,
    minimum height=2.4em]
\tikzstyle{plate} = [draw, rectangle, text width=6.2cm, 
    text centered, minimum height=15em]
\tikzstyle{ghost} = [draw=none,ellipse, text width=2.8em,
    minimum height=2.4em]
    
\slugcomment{Submitted to the Astrophysical Journal on September 9, 2014}
\shorttitle{Composition Distribution of Sub-Neptunes}
\shortauthors{Wolfgang \& Lopez}

\begin{document}

\title{How Rocky Are They?  The Composition Distribution of \emph{Kepler}'s Sub-Neptune Planet Candidates within 0.15 AU}

\author{Angie Wolfgang\altaffilmark{1} and Eric Lopez}
\affil{Department of Astronomy and Astrophysics, \\University of California,
    Santa Cruz, CA 95064}
\altaffiltext{1}{NSF Graduate Research Fellow}
\email{wolfgang@ucolick.org}

\begin{abstract}

The \emph{Kepler Mission} has found thousands of planetary candidates with radii between 1 and 4 R$_\oplus$.  These planets have no analogues in our own Solar System, providing an unprecedented opportunity to understand the range and distribution of planetary compositions allowed by planet formation and evolution.  A precise mass measurement is usually required to constrain the possible composition of an individual super-Earth-sized planet, but these measurements are difficult and expensive to make for the majority of \emph{Kepler} planet candidates.  Fortunately, adopting a statistical approach helps us to address this question without them.  In particular, we apply hierarchical Bayesian modeling to a subsample of \emph{Kepler} planet candidates that is complete for $P< 25$ days and $R_{pl}>1.2$ R$_\oplus$ and draw upon interior structure models which yield radii largely independent of mass by accounting for the thermal evolution of a gaseous envelope around a rocky core.  Assuming the envelope is dominated by hydrogen and helium, we present the current-day composition distribution of the sub-Neptune-sized planet population and find that H+He envelopes are most likely to be $\sim 1\%$ of these planets' total mass with an intrinsic scatter of $\pm 0.5$ dex.  We address the gaseous/rocky transition and illustrate how our results do not result in a one-to-one relationship between mass and radius for this sub-Neptune population; accordingly, dynamical studies which wish to use \emph{Kepler} data must adopt a probabilistic approach to accurately represent the range of possible masses at a given radius.

\end{abstract}


\section{Introduction} \label{intro}

The \emph{Kepler Mission} has found thousands of planetary candidates with sizes between that of Earth and Neptune \citep{Rowe14,Bur14,Bat13,Bor11}.  Considering that no such planets exist in our own Solar System, this discovery elicits fundamental questions about their nature: are these planets scaled-up versions of Earth, scaled-down and irradiated versions of Neptune, or something in-between?  What is the ``typical" composition of this planet population, and what is the range of possibilities, as constrained by the planets we have observed?  At what radius is the expected transition between rocky and gaseous compositions?

Addressing such population-wide inquiries about bulk compositions requires two tools: first, models of internal structures which relate an individual planet's composition to its measured radius, and second, a statistical framework which combines information about individual members of a population into an inference about the whole while appropriately accounting for uncertainties in individual observables.  The former has been studied by a number of authors, as summarized below; the latter, however, has received limited treatment in the exoplanet literature.  Here we provide an exoplanet-specific example of one such statistical tool commonly used for population studies in other fields, and in doing so answer questions about the range and distribution of compositions for these sub-Neptune-sized planetary candidates.

\subsection{Modeling Sub-Neptune Interior Structures} \label{intromodels}

Theoretical modeling of exoplanet interiors has a substantial history, starting with models that were developed to understand the structure and evolution of gas giants \citep[e.g.][]{For07,Mar07}.  As recent surveys have uncovered ever smaller extrasolar planets, these models have been extended to the new population of sub-Neptune-sized planets.  Studies of such low-mass planets include investigations of ``ocean worlds" \citep{Leg04}, low-density irradiated exo-Neptunes \citep{Rog11}, and scaling relations between mass and radius for sub-Jovian planets of varying compositions, including iron, silicates, water ice, carbon compounds, hydrogen/helium, and various combinations thereof \citep{Val06,Sea07,For07}.

These models have been applied to numerous individual Neptunes and sub-Neptunes, such as GJ 876d \citep{Val07a}, CoRoT-7b \citep[e.g.][]{Leg09,Val10,Jac10}, GJ1214b \citep[e.g.][]{Cha09,Rog10b,Net11,Val13}, and the Kepler-11 system \citep{Lop12}, to provide constraints on what their bulk compositions could be.  Inferring more detailed composition parameters like an individual planet's core, mantle, and envelope mass is of course a highly degenerate problem made worse by the possible choices for the number and type of layers in the planet's interior \citep{Val07b,Rog10a}.  Nevertheless, we can derive some guidance for how to best address this problem and make progress on answering these population-wide composition questions by noting a few salient characteristics of the overall low-mass planet population.  First, a substantial fraction of these planets have radii that are just too large to be explained by rock/ice/water combinations \citep[e.g,][]{Lop14,Rog14}.  Second, mass constraints for a few dozen sub-Neptunes indicate that planets at the same radii can vary in mass by a factor of $\sim 2-4$ \citep{Mar14, Wei14}, hinting at significant compositional variability within this population.  Finally, conclusively rocky compositional constraints have been obtained for a few small planets, most notably CoRoT-7b, Kepler-10b \citep{Bat11}, and Kepler-78b \citep{Pepe13,How13}.

Based these observations, we adopt a few key assumptions which allow us to move forward with this work, to infer compositions for \emph{Kepler}'s sub-Neptune population.  First, knowing that some of these planets are decisively rocky while others must contain non-negligible amounts of hydrogen and helium motivates the assumption of a rocky core with a H+He envelope for all of these low-mass planets.  Of course, this does not mean that ``water worlds", i.e. planets with either a distinct water layer or with water vapor comprising a substantial percentage of the gaseous envelope, could not exist.  Indeed, if photoevaporation plays a significant role in shaping this irradiated planet population, planets which are shown to exclusively lie in the radius-flux ``occurrence valley" predicted by \citet{Lop13} and \citet{Owen13} are likely such water worlds.  Nevertheless, with core accretion as a reasonable proposal for the formation of these sub-Neptunes and with protoplanetary disks composed primarily of hydrogen, the most straightforward explanation for the substantial compositional variation implied by measured masses and radii is variation in the accretion and loss of hydrogen, rather than the somewhat extreme dynamic range in the processes of ice differentiation, disk migration, and water evaporation needed to produce an entire population of water worlds which match these observations.  While both of these ideas merit further work and observational testing, Occam's Razor
drives us to assume a rock plus hydrogen composition for this study.

With the postulate that gaseous envelopes tend to dominate the non-rocky portion of the planet's structure, we can adopt a two-component interior structure and, for now, set aside the problem of compositional degeneracy that arises from models with three or more layers.  Even so, a large amount of theoretical uncertainty in the intrinsic luminosity of these planets remain.  Fortunately, coupling interior structure models to atmospheric radiative transfer models \citep[e.g.][]{For07,Gui10} enables tracking of the thermal cooling of a planet's interior as it ages, eliminating the need to marginalize over the internal energy \citep{Lop12,Lop14}. When applied to highly irradiated sub-Neptunes as done in \citet{Lop14}, these thermally evolving models result in radii that are more sensitive to the fraction of a planet's mass that is in a hydrogen and helium envelope than to the total mass.  This has significant implications, as mass measurements are not needed to get a sense for this composition parameter, and the information content in the \emph{Kepler} radius distribution can be maximally leveraged for such studies.

\subsection{Statistical Treatment of Planet Populations}

Radial velocity (RV) surveys provided the first opportunity to study the characteristics of an emerging population of planets.  \citet{Tab02} and \citet{Cum08} used a maximum likelihood approach with Poisson statistics to infer the joint mass-period distribution of detected RV planets, while necessarily incorporating RV detection thresholds to account for incompleteness (known as ``truncation" in the statistical literature).  The eccentricity distribution of RV planets were studied by \citet{Jur08} and \citet{Ford08}, among others, who performed K-S tests to compare the eccentricity distributions of their planet-planet scattering simulations to various sub-populations of RV planets to illuminate their origins.  More recently, \citet{How10} and \citet{May11} extended the analysis of survey incompleteness to smaller masses, using a simple efficiency correction to obtain binned estimates of the occurrence rate of RV planets.

These population studies have expanded in scope and feasibility with the advent of the \emph{Kepler Mission} \citep{Bor10, Koc10}, which discovered over 3500 planet candidates in its first three years of data (\citealt{Rowe14}; see \citealt{Bur14} for the latest published planet candidate catalog).  Among them are analyses of the occurrence rate of exoplanets at different sizes and periods \citep{You11, How12, Dong13, Fre13, Dre13, Mor13, Pet13}, the consistency between RV and transit surveys \citep{Wol12, Wri12, Fig12}, the \emph{Kepler} eccentricity distribution \citep{Moo11, Kane12, WuY13}, and the frequency of multiple-planet systems \citep{Lis11, Tre12, Fab12}.  The relationship between the radii and masses for sub-Neptune planets has also received attention \citep{Wol12,WuY13,Wei14}, with the aim of illuminating the compositions of these planets.

These studies use a range of statistical techniques, including the intuitive yet idealized inverse efficiency method \citep{How12,Pet13}; linear regression on binned, mean estimates of very uncertain, intrinsically dispersed individual points \citep{Wei14}; Monte Carlo approaches \citep{Wol12,Fre13}; maximum likelihood that incorporates survey incompleteness \citep{You11,Tre12,Dong13}; and non-parametric kernel density estimation \citep{Mor13}.  Each of these studies treat error in the observed quantities of individual planets differently, but none incorporate them in a way that produces rigorous posterior estimates of the population parameters of interest.  Given that the goal of such population studies is to characterize the population given the observed data, the quality of this data should play a large role in the inference of the population parameters.  

Hierarchical Bayesian modeling (HBM) is very naturally suited to this problem, and has been in use for decades by many fields, including bioinformatics and political science, whose key science questions involve inferring characteristics of a population from noisy observations of individual members.  The HBM framework is very general, and its usefulness extends to a number of commonly encountered problems in astronomy; we refer the reader to \citet{Lor07} and \citet{Lor12} for a discussion of multi-level modeling in a general astronomical context, and to \S \ref{method} for an overview of its capabilities for the science goals of this work.  

The promise that HBM holds for exoplanet population studies has only recently been realized.  The first instance of HBM in the exoplanet literature is \citet{Hog10}; they derive an importance sampling algorithm which incorporates posterior samples that have already been computed for individual planets into inferences about the population, and they apply this algorithm to the planetary eccentricity distribution.  \citet{FoM14} and \citet{Rog14} use this algorithm to infer the occurrence rate of planets as a function of period and radius, and to infer the radius at which super-Earths transition from gaseous to rocky compositions, respectively.  This work, on the other hand, is the first study to perform full hierarchical Bayesian modeling where inferences on both the population and the individuals are made; the details are laid out in \S \ref{ourHBM} - \ref{individcomp}.

In this paper we present the first quantitative distribution of sub-Neptune compositions, which we define as the fraction of a planet's mass that exists in a hydrogen and helium envelope around an Earth-like rocky core that can vary in mass.  In \S \ref{Kepdata}, we describe the \emph{Kepler} planet candidates and how the sample used for this work was selected.  In \S \ref{method} we explain what hierarchical Bayesian modeling is and detail the specifics of the model that we use to obtain the sub-Neptune composition distribution presented in \S \ref{Res}.  We discuss the implications of these results in \S \ref{Discuss}, and conclude in \S \ref{Conclu}.

\section{\emph{Kepler} Objects of Interest} \label{Kepdata}

During its regular mission, \emph{Kepler} stared at nearly 200,000 stars to detect transiting extrasolar planets.
The search for transit signatures among such a large target sample requires a substantial amount of processing, both by computers and by humans, to produce the list of high fidelity planetary candidates that we use in this work.  The automated portion of this process is the pipeline developed by the \emph{Kepler Mission} Science Operations Center (SOC), as outlined in \citet{Jen10a}; it includes modules that calibrate the raw images from the \emph{Kepler} spacecraft \citep{Qui10}, produce raw photometry from these images \citep{Twi10b}, detrend the raw photometric time series to remove systematic instrumental effects \citep{Twi10a}, search for transit-like signals using a wavelet-based adaptive-matched filter \citep{Jen10b}, and then fit a transit model to further characterize the signal \citep{WuH10}.

The human effort continues where this automated pipeline ends: with the list of Threshold Crossing Events (TCEs), the periodic transit-like signals which pass the pipeline's filter.  The number of TCEs can be substantial, on the order of $10^4$ \citep{Ten13,Ten14}, and a large number are produced by astrophysical variability or instrumental noise ($> 80$\% of Q1-8 TCEs;  \citealt{Bur14}).  To identify these cases, the TCE list undergoes a ``triage" stage where the Data Validation summary pages produced by the last module in the SOC pipeline are visually inspected by members of the Threshold Crossing Event Review Team (TCERT).  If the shape of the signal has the characteristic U- or V-shape of a planetary transit or stellar eclipse, it is made into a \emph{Kepler} Object of Interest (KOI) and passed onto the next stage of human inspection where it is given either a false positive or a planetary candidate disposition.

The data and metrics used to disposition KOIs are explained in detail in \citet{Bat13}, \citet{Bur14}, and \citet{Rowe14}.  For our purposes, it is sufficient to note that this determination only considers data obtained by the \emph{Kepler} spacecraft, and that an ``innocent until proven guilty" approach is adopted.  Additional information in the \emph{Kepler} light curves and pixel images does enable identification of some astrophysical false positives, i.e. unequal-mass eclipsing binary systems or background eclipsing binaries with a significant sky-projected offset from the target star.  However, the majority of KOIs lack the unambiguous evidence of these false positive scenarios or need additional observations to provide this evidence, and therefore are determined to be (or ``dispositioned" as) planetary candidates (hitherto ``PCs").  In general, what counts as unambiguous evidence is strictly defined (i.e. a large transit depth or a V-shaped transit is not considered sufficient evidence for a false positive disposition), which results in some KOIs being optimistically labeled as PCs.

\subsection{KOI False Positives and Incompleteness}\label{incomplete}

A number of studies have analyzed the effect of this inclusive approach on the reliability of the PC catalog.  While \citet{Mor11} found that overall the false positive probability (FPP) of KOIs labeled as planetary candidates are low ($\sim 5$\%), this FPP does increase to $\sim 10$\% for PCs with larger radii.  \citet{San12} qualitatively corroborates this effect via the radial velocity technique by measuring masses of the detected transiting companions, reporting that 35\% of the deepest, shortest period PC signals have masses too high for planetary bodies.  The analysis of \citet{Fre13} also supports this finding with a more detailed treatment of the unknown planet occurrence rate, finding FPP above 15\% for PCs with radii $> 4 $R$_\oplus$ as opposed to FPP $\sim 8$\% for the smaller Super-Earths.

The converse issue of completeness is also necessary to address when using the \emph{Kepler} PC catalog for statistical studies.  There are a number of reasons why the catalog does not necessarily reflect the intrinsic distribution of planet properties, and these effects must be properly accounted for before one can make any conclusions about the general characteristics of these planets.  The most obvious of these is the transit probability: because a planet is more likely to transit its host star when the radius of the star is larger and the planet-star distance is smaller, \emph{Kepler} will naturally detect more short period planets and planets around smaller stars.  This effect must be corrected if the planet property of interest is expected to correlate with either stellar type or planet period.  

\begin{figure*}[t]
\begin{center}
\includegraphics[angle=270, scale=0.67]{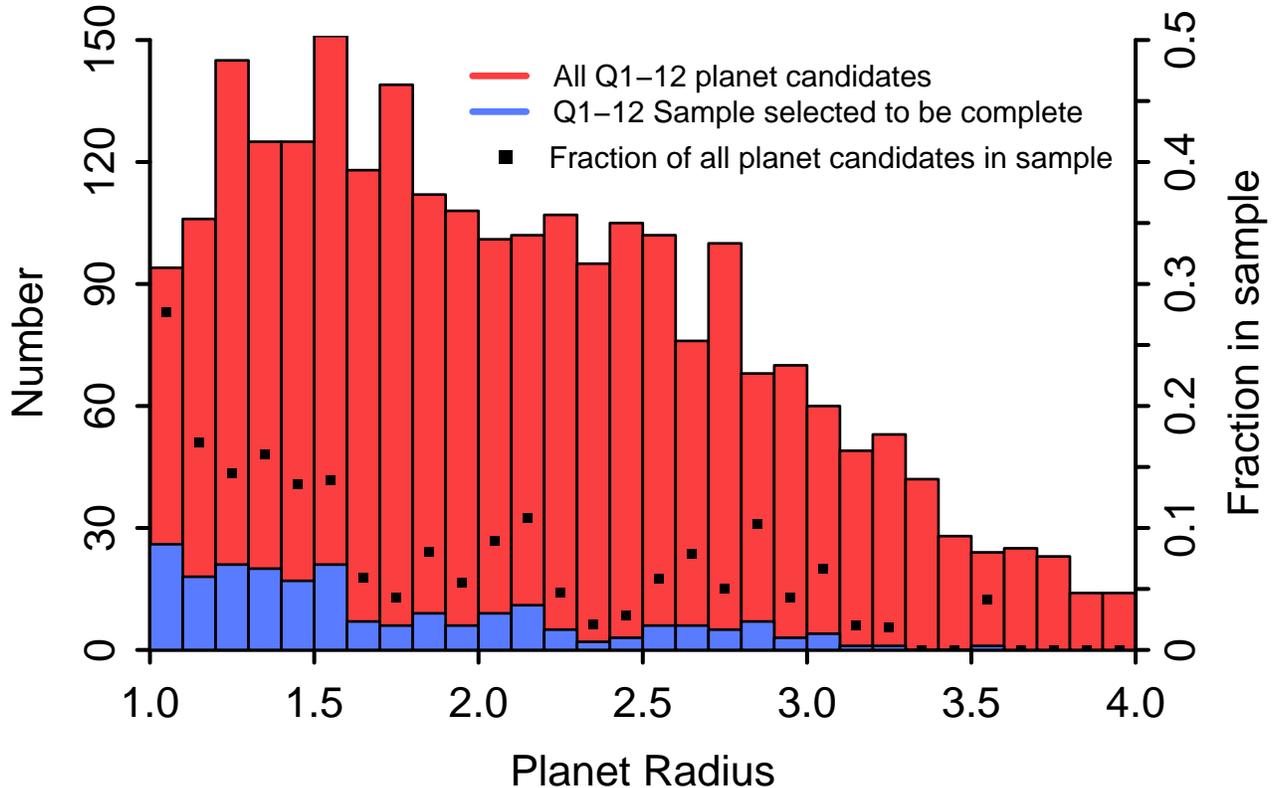}
\caption{Radius distribution of our subsample (blue; $N= 215$) of sub-Neptune planet candidates (PCs) compared with the total distribution of Q1-12 PCs in this size range (red; $N=2572$).  Detection biases cause fractionally fewer small planets to be found, especially at longer periods.  Carefully restricting the parent star sample and imposing a period cut, as was done to create the subsample, can mitigate these biases (see \S \ref{sampsel} for details).  The black points corresponding to the right y-axis quantify this mitigation, showing the fraction of PCs in that radius bin from the total Q1-12 catalog which made it into our more complete subsample.} \label{samp}
\end{center}
\end{figure*}

In addition, \emph{Kepler} is not able to detect every transiting planet in its field of view, as the stellar noise profile varies strongly from star to star.  Given the range of noise levels across the target star sample, this leads to detecting a lower fraction of existing transiting planets at smaller radii and longer periods.  Fortunately, limiting the considered target star sample to the least noisy stars, as defined by the Combined Differential Photometric Precision (CDPP) calculated by the \emph{Kepler} pipeline \citep{Chr12}, can diminish this effect.  

Finally, \citet{Bat13} showed that the \emph{Kepler} pipeline outlined above detects fewer low signal-to-noise transit-like signals than expected, pointing to yet another detection bias against small, long-period planets.  This pipeline incompleteness has only been recently realized to have a significant effect, and work to characterize it via transit injection is ongoing.  First results presented in \citet{Chr13} indicate that the later modules of the pipeline do not systematically perturb the signal strength of individual events in an individual quarter of data, although larger perturbations are measured for lower signal-to-noise events.  On the other hand, studies over multiple quarters shows that the transit search module culls the lower signal-to-noise events to a significant degree, due in large part to metrics that have been implemented to discard false alarm detections (private communication, \emph{Kepler} Completeness Working Group).  This aggressive rejection results in significant incompleteness ($< 95\%$) for transits with a phased and folded signal-to-noise ratio (SNR) $< 15$.

\subsection{Selecting a Complete Subsample} \label{sampsel}

To select our sample, we begin with the cumulative \emph{Kepler} Objects of Interest (KOI) table available at the NASA Exoplanet Archive \citep{Ake13}, which at the time of access (December 2, 2013) consisted of the Q1-12 catalog \citep{Rowe14}, a heterogenous list of KOIs identified in the first 12 quarters or less of \emph{Kepler} data.  The heterogeneity arises from the fact that, in general, the higher signal-to-noise (S/N) events are identified with fewer data and a less mature vetting process (overviewed in \S \ref{Kepdata}).  Furthermore, the reported planet parameters in the cumulative catalog are derived from different total amounts of data.  

Even though this list does not yet represent the uniform sample that is ideal for statistical studies, it is the best-knowledge catalog to date.  Uniformity in planet parameters, if not in the planet candidate (PC) disposition itself, can be improved by matching the KOIs to the latest Threshold Crossing Events (TCEs) (\citealt{Ten14}; see \S \ref{Kepdata} for a discussion of the differences between these lists; vetting for the Q1-16 planet candidate catalog had not yet started at the time of this analysis).  Matching Q1-12 PCs to Q1-16 TCEs also ensures that the planet parameters used in our analysis are those derived with the best-knowledge stellar parameters: the Q1-16 stellar properties are described in \citet{Hub14}, hereafter referred to as Hub14.  

Starting with the 3601 KOIs listed as PCs in the Q1-12 catalog, we retain 3322 PCs whose stars have a Q1-16 TCE within 1\% of the PC period $P$ and an epoch modulo $P$ within $0.05*P$ days of the PC epoch.  Half of the discarded PCs do not have any Q1-16 TCEs identified for that target star.  This could be due to the pipeline's mistaken removal of short-period, high-S/N transits via the narrow-band oscillation filter described in \citet{Ten14}, or to strong transit timing variations (TTVs), or to PCs that are actually false alarm detections.  The other half of the discarded PCs with non-matching periods and epochs can also be explained by TTVs or false alarms, or by the more common circumstance where the pipeline identifies a harmonic or sub-harmonic of the true transit signal \citep{Ten14}.

In this work we characterize the compositions of sub-Neptune planets, so we limit ourselves to the 2572 PCs with 1 R$_\oplus < R_{pl} < 4$ R$_\oplus$.  Because we are using the Q1-16 TCE parameters, these radii are derived from the Q1-16 data using the Q1-16 stellar parameters of Hub14.  The remainder of our sample cuts arise from concerns about the completeness of this sub-Neptune sample.  Because our analysis method (overviewed in \S \ref{method}, detailed in \S \ref{ourHBM}) automatically folds the shape of the PC radius distribution into our result on the composition distribution of sub-Neptunes, we must take precautions to ensure that this PC radius distribution is as close to the true planet radius distribution as possible.  Figure \ref{samp} shows the radius distribution of our final sample compared with the total distribution of Q1-12 PCs.  Per the discussion in \S \ref{incomplete}, we expect that the full sample is less complete at smaller radii; the black points, which denote the effective completeness correction made by choosing a complete subsample, illustrate that this is indeed the case.  Thus,  the cuts described below are effective in minimizing the detection biases present in the larger catalog.

Using the standard S/N calculation for a transiting planet (see, for example, \citealt{Wol12}) and \emph{Kepler}'s detection criteria of 7.1$\sigma$, we find that a $R_\star < 1.2$ R$_\odot$ star with noise $< 100$ ppm on transit duration timescales that had been observed continuously for three years should be complete for planets with $P < 25$ days and $R_{pl} > 1.2$ R$_\oplus$.  This radius cut encompasses the vast majority of planets which are conservatively expected to still have a gaseous envelope, and so preserves completeness of the planets which contribute to our composition distribution.  We therefore restrict our sample to main-sequence host stars (log($g$) $> 4.0$) with $R_\star < 1.2$ R$_\odot$ and a CDPP value that when scaled to the duration of the planet's transit is less than 100 ppm.  We further require that the host star have been observed for all 12 quarters.   With the final cut on period, we retain a sample size of 215 sub-Neptune sized planets within $\sim 0.15$ AU of their host stars. 

\begin{figure}[t]
\begin{center}
\includegraphics[angle=270, scale=0.3]{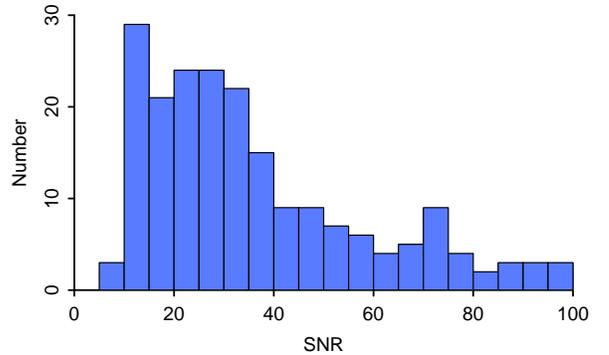}
\caption{Distribution of the Q1-16 signal-to-noise ratios of the majority of our sample, scaled to the Q1-12 detection baseline (14 PCs have scaled SNR $> 100$ and are not displayed).  32 PCs have SNR $< 15$, where pipeline incompleteness becomes significant; of these, two have $R_{pl} >1.6$ R$_\oplus$.  Because only $\sim 1$\% of our sample suffers from pipeline incompleteness while having nonzero gaseous envelope mass fractions, our composition distribution (\S \ref{compdist}) is not affected by this otherwise problematic sample bias.} \label{snrhist}
\end{center}
\end{figure}

Although this careful selection of the host star sample accounts for detection bias, pipeline incompleteness is still a concern.  To assess how much of an effect this could have on our results, we scale the Q1-16 transit model SNRs of our Q1-12 sample to the time baseline over which they were detected, and display the results in Figure \ref{snrhist}.  We note that only 15\% of our sample has SNR $< 15$, where pipeline incompleteness becomes significant; of these, 95\% have $R_{pl} < 1.6$ R$_\oplus$, which we show in \S \ref{rockgastrans} are most probably rocky given our ``best-fit" composition distribution.  Therefore, pipeline incompleteness does not affect the distribution of gaseous mass fractions that we infer from \emph{Kepler}'s irradiated sub-Neptune population (\S \ref{compdist}).

\section{Methods: Characterizing Planet Compositions via Statistical Modeling} \label{method}

The goal of this work is to understand the range of gaseous envelope mass fractions that \emph{Kepler}'s super-Earths and sub-Neptunes can possess.  In this section, we motivate why hierarchical Bayesian modeling (HBM) is such a natural approach to this problem, discuss some of its advantages over other methods, and detail the specific model that we use to infer the sub-Neptune composition distribution.  Due to the limited use of HBM in the exoplanet literature we spend significant time explaining the reasons, context, and application of this choice, but for the hurried reader we provide the following summary:

\begin{itemize}[leftmargin=4mm]
    \item Hierarchical Bayesian modeling is the natural choice for constraining the population distributions of exoplanet properties (such as compositions or radii), when those properties are either unobserved (compositions) or possess significant errors (radii).
    \item HBM is also the natural choice when the priors on individual exoplanet properties are expected to have an intrinsic scatter instead of one true value, where the scatter is due to some physical variation among the population and is of scientific interest.
    \item HBM provides posteriors on both the population parameters (e.g., the mean of the composition distribution) and on the individual parameters (e.g. an individual planet's composition), thereby enabling simultaneous inference on individual planets and the population as a whole.
    \item By relating individuals to each other through this hierarchical framework, HBM provides posterior estimates of individual exoplanet properties which have smaller variance than if multiple individual Bayesian analyses were performed independently.  This is called ``shrinkage" and is illustrated in \S \ref{convshrink}.
    \item HBM is a straightforward extension of regular Bayesian modeling, requiring only the definition of conditional probability and a slight shift in interpretation, and so uses the same basic computational algorithms such as Markov Chain Monte Carlo.
    \item As with all Bayesian analysis, HBM enables prediction of future observations by marginalizing the likelihood over the posterior distributions.  We present the sub-Neptune posterior predictive composition distribution in Figure \ref{fenvpost}.
\end{itemize}

The discussion below details the application of hierarchical, or multi-level, modeling to exoplanet compositions; for a more general discussion of the past use and future promise of multi-level modeling in astronomy, we refer the reader to \citet{Lor07, Lor12}.

\subsection{Choosing an Appropriate Statistical Framework}\label{choosing}

\begin{figure}[h]
\begin{center}
\begin{tikzpicture}[node distance = 1.5cm, auto]
    \node [param, fill=yellow!50] (fenv) {$f_{env}$};
    \node [param, left of=fenv, node distance = 2cm] (mpl) {$M_{pl}$};
    \node [param, right of=fenv, node distance = 2cm] (flux) {$F$};
    \node [param, right of=flux, node distance = 2cm] (age) {$t_{pl}$};
    \node [obsable, above of=fenv] (per) {$P$};
    \node [param, right of=per, node distance = 2cm] (teff) {$T_{eff}$};
    \node [param, right of=teff, node distance = 2cm] (mstar) {$M_\star$};
    \node [param, below of=fenv] (rpl) {$R_{pl}$};
    \node [param, right of=rpl, node distance = 2cm] (rstar) {$R_\star$};
    \node [param, left of=rpl, node distance = 2cm] (b) {$b$};
    \node [obsable, below of = rpl] (depth) {$\delta$}; 
    \path [line] (fenv) -- (rpl);
    \path [line,dashed] (mpl) -- (rpl);
    \path [line,dashed] (flux) -- (rpl);
    \path [line,dashed] (age) -- (rpl);
    \path [line] (per) -- (flux);
    \path [line] (teff) -- (flux);
    \path [line,dashed] (mstar) -- (flux);
    \path [line] (rpl) -- (depth);
    \path [line] (rstar) -- (depth);
    \path [line,dashed] (b) -- (depth);
\end{tikzpicture}  \vspace{4mm}
\caption[blah]{
The relationships between the quantities that \emph{Kepler} observes (gray ellipses) and the quantity of interest in this work (highlighted in yellow).  This diagram represents the flow of information for a single planet candidate.  First-order relationships, i.e. those that dominate the value of the resulting quantity, are denoted as solid lines whereas second-order relationships are represented by dashed lines.  Note that the mapping from the second line to the planet radius is given by the models of Lop14.  These quantities are defined as follows: \\
    \begin{itemize}[label={}]
       \item $P$ = period 
       \item $T_{eff}$ = effective temperature of host star
       \item $M_\star$ = mass of host star
       \item $M_{pl}$ = total mass of planet
       \item $f_{env}$ = fraction of $M_{pl}$ existing in a gaseous H+He envelope
       \item $F$ = stellar flux incident on the planet
       \item $t_{pl}$ = age of planet
       \item $b$ = impact parameter
       \item $R_{pl}$ = radius of planet
       \item $R_\star$ = radius of host star
       \item $\delta$ = transit depth 
    \end{itemize}
} \vspace{-4mm} \label{basicstruct}
\end{center}
\end{figure}
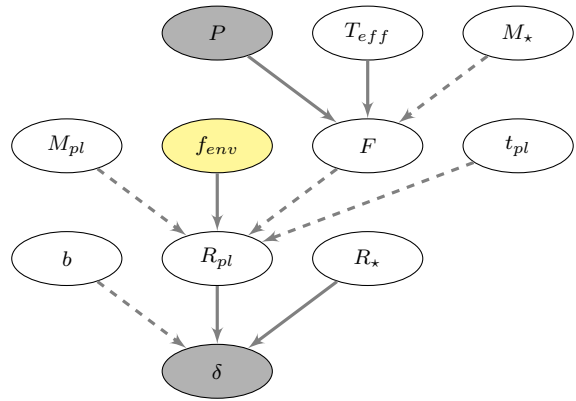

To understand why we have chosen HBM to solve this problem, we must first understand how these planets' compositions relate to the quantities that \emph{Kepler} measures.  Most simply, a sub-Neptune's gaseous mass fraction sets its radius, as Lop14 showed that these planets' compositions dominate over other factors in determining their size; the radius, in turn, is primarily derived from the depth of the transit signal, which is the quantity that \emph{Kepler} directly observes.  In practice, however, several other quantities become important to include in order to accurately infer planetary compositions from their transit parameters; the relationships between them for a single planet candidate are shown in Figure \ref{basicstruct}.

The hierarchical structure of this problem is immediately apparent.  Having such a multi-tiered relationship between relevant quantities does not necessarily require a hierarchical Bayesian framework, however.  Simple inversion of the problem and standard error analysis is sufficient if the relationships are deterministic (that is, they can be summarized as a function that maps one set of input values onto one output value) and if the values of the quantities themselves are well known with errors that are either small or well-behaved (i.e. symmetric and uncorrelated).

For the problem outlined in Figure \ref{basicstruct}, the relationships could indeed be deterministic (but see the discussion about likelihoods below and the full problem outlined in Figure \ref{fullstruct} and Equations (\ref{fulleqn}) - (\ref{statmod})).  On the other hand, the values of many of the quantities in Figure \ref{basicstruct} are neither well known nor are their uncertainties well-behaved.  For example, the \emph{Kepler} pipeline described in \S \ref{Kepdata} produces biased and poorly constrained estimates for impact parameter $b$ (Rowe, private communication).  Even more problematic is the dependence of these quantities on stellar parameters: Hub14 illustrates that the current state of the observations of \emph{Kepler}'s target stars leads to large, asymmetric uncertainties on $R_\star$ and $M_\star$, and can even make them multimodal (see Figure 8 of that paper).  As a result, the envelope fractions $f_{env}$ cannot be straightforwardly calculated from the observed transit depths $\delta$ under the assumption of small errors, and a more sophisticated analysis is warranted.  

To incorporate our observational uncertainty, we must relax the requirement of deterministic relationships and allow probabilistic, or stochastic, relationships within the structure of Figure \ref{basicstruct}.  This is accomplished by computing the likelihood function, which describes how probable the data are under their measurement uncertainty, given different values for the model parameters.  Choosing the appropriate likelihood function requires knowledge about how the measurement errors behave; often it is assumed that they follow a Gaussian distribution, meaning that the measured values are normally distributed around the true value.  Answering the question of interest, i.e. ``what is the gaseous envelope mass fraction of planet X", then involves inference, where one identifies the parameter values which ``best fit" the data.  

``Best fit" parameter values can be found by maximizing the likelihood function directly, which gives an estimate of the ``true" value exhibited by nature.  Alternatively, by shifting one's interpretation of the likelihood to allow for uncertainty in the true parameter values, one can combine the likelihood with some prior information to create a posterior distribution of likely parameter values.  In practice, the former method of maximum likelihood (ML) often manifests as calculating $\chi^2$, which requires the aforementioned assumption of normally distributed errors (note, however, that ML can be performed for any arbitrary likelihood function, by solving for the parameter values at which such a likelihood is maximized).  In contrast, the latter method of Bayesian inference usually involves Markov Chain Monte Carlo (MCMC) simulations, wherein a sequence of posterior probabilities is numerically computed in a way that optimally explores the range of parameter values allowed by the data.

While the choice between using ML and Bayesian inference is often a matter of philosophical preference (see \citealt{Lor12} for an in-depth discussion on the philosophical differences between frequentist and Bayesian approaches), ML has the strongest computational advantage over Bayesian methods when one has no prior information and when the likelihoods are easy to write down, analytically tractable, and do not involve too many parameters.  In that case, the matrix inversion required to find the best-fit values can be performed quickly and efficiently, and the confidence intervals in those best-fit values can be analytically computed.  

Unfortunately, such an analytic treatment is not possible for the problem we endeavor to solve in Figure \ref{basicstruct}, given the necessarily numerical calculation of the stellar parameters and their errors, which are non-Gaussian.  In such a situation, ML would involve computing likelihoods on a grid of parameter values, and error bars would be interpreted as the range of parameter values which enclose the maximum likelihood estimate for 68\% of the datasets.  Not only is the latter task difficult to do with a single dataset, but this approach is much less computationally efficient than a Bayesian treatment involving MCMC, as the Markov Chain spends less time exploring parameter space that has low probability of matching the data.  This computational consideration, in combination with the realization that a Bayesian approach is better suited to our problem, where we only have a single list of planet detections from \emph{Kepler} and significant uncertainty about the true physical parameters of the planet population, guides our choice of a Bayesian framework for this study.  In doing so, we also enable the incorporation of prior information, which can naturally be extended into the hierarchical structure appropriate for this problem, as explained in \S \ref{HBM}.

\subsection{Applying Bayes' Theorem}\label{basicBayes}

The basic Bayesian framework is readily summarized with the following interpretation of Bayes' Theorem: \\
\begin{align}  \label{bayes}
p(\bm{\theta}|\bm{X}) = \frac{p(\bm{X}|\bm{\theta})p(\bm{\theta})}{p(\bm{X})} ,
\end{align} \\ 
where $\bm{\theta}$ is the set of parameters that define the model (i.e. the quantities circumscribed by ellipses in Figure \ref{basicstruct}), $\bm{X}$ is the set of data values (i.e. the quantities circumscribed by rectangles in Figure \ref{basicstruct}), and $p(x|y)$ denotes the probability distribution function of quantity x at a given value of quantity y (in other words, the probability of x conditional on y).  Inference occurs via the posterior distribution $p(\bm{\theta}|\bm{X})$, which yields the probability of various parameter values given the data; the ``best fit" values can be the mode, median, or some other central statistic of this distribution.  Computing the posterior requires specifying the likelihood $p(\bm{X}|\bm{\theta})$ as described above and the prior distribution $p(\bm{\theta})$, which reflects previous information about how intrinsically likely different parameter values are; the normalizing constant $p(\bm{X})$ can be ignored when one uses MCMC to compute the posterior numerically, as the core of the MCMC algorithm involves computing posterior probability ratios within which this constant cancels.

To apply this framework to our problem, we note that transit depth $\delta$ is our primary observable quantity, so we set $\bm{X} = \{\delta\}$.  $\bm{\theta}$ therefore denotes the rest of the unobserved quantities in Figure \ref{basicstruct}.  A full Bayesian treatment would require specifying the joint prior probability function $p(R_{pl},R_\star,b,M_{pl},f_{env},F,T_{eff},M_\star,t_{pl})$ along with our likelihood $p(\delta|R_{pl},R_\star,b,M_{pl},f_{env},F,T_{eff},M_\star,t_{pl})$, but in practice the varying levels of importance in the relationships between the different quantities, as denoted by the dashed vs. solid lines in Figure \ref{basicstruct}, allow us to simplify the problem.  Accordingly, we hold constant the values of parameters that are related to a second-order quantity, thereby setting $P$ to the observed value and $T_{eff}$ and $M_\star$ to the best-fit values determined by Hub14.  We also set to fiducial values second-order parameters such as $b$($=0$) and $t_{pl}$($=5$Gyr) that are not well constrained by the data.

With these modifications, our prior probability has simplified to $p(R_{pl},R_\star,M_{pl},f_{env},F)$, but still has a nontrivial functional form given the hierarchical dependence between the parameters.  Fortunately, we can put this intrinsic structure to use: rather than specify one distribution containing all of these parameters, we instead use the definition of conditional probability to derive a joint prior probability distribution in terms of conditional distributions.  This definition states that
\begin{align}
p(x,y) = p(x|y)p(y) \label{condprob} ,
\end{align}
where $p(x,y)$ is the joint probability distribution of x and y (i.e. it states the probability of both of those x and y values occurring), $p(x|y)$ is the conditional probability distribution of x given y (i.e. at a set value of y, it states the probability of an x value), and $p(y)=\int p(x,y)dx$ is the marginal probability distribution of y (i.e. it states the probability of the y value occurring under all conditions).  Therefore, we can split our joint prior probability distribution into a series of conditional and marginal distributions as appropriate given the structure of our problem: \\
\begin{align} \label{priorstruct}
& p(R_{pl},R_\star,M_{pl},f_{env},F) \nonumber \\
  & \qquad = p(R_{pl}|R_\star,M_{pl},f_{env},F) \\
  & \qquad \qquad \times p(R_\star|M_{pl},f_{env},F)p(M_{pl},f_{env},F) \nonumber \\
 & \qquad = p(R_{pl}|M_{pl},f_{env},F)p(R_\star)p(M_{pl})p(f_{env})p(F) .  \nonumber
\end{align} \\
Note that in simplifying the right-hand side we have assumed that $M_{pl}$, $f_{env}$, and $F$ are independent of each other, that $R_\star$ is independent of $M_{pl}$, $f_{env}$, and $F$, and that the true, intrinsic radius of the planet $R_{pl}$ is independent of $R_\star$; this is also reflected in the structure of Figure \ref{basicstruct}.  The usefulness of this framework is that such dependencies can be effortlessly included in subsequent analysis should there be good reason to expect that they exist or are important.

To make any further progress, we must specify what functional forms these prior probabilities should take.  $p(R_\star)$ is the distribution of allowed radius values for the host star, and is equivalent to the likelihood numerically calculated by Hub14 after it has been marginalized over all other stellar parameters; inclusion of this distribution among our prior information is how we are able to account for uncertainties in the stellar parameters.  $p(R_{pl}|M_{pl},f_{env},F)$ represents the Lop14 sub-Neptune internal structure models.  Because these models map a planet's mass, envelope fraction, and incident flux to a single radius value, this probability distribution is actually a delta function; this is how we allow for deterministic relationships in our probabilistic model.  Due to the simplifying choices we made above, we have also forced $p(F)$ to be a delta function.  Implicit marginalization over these last two parameters with delta function probability distributions then allows us to write: \\
\begin{align}
p(R_{pl},R_\star,M_{pl},f_{env},F) = p(R_\star)p(M_{pl})p(f_{env}) ,
\end{align} \\
where $R_{pl}$ will show up in the likelihood as a deterministic function of $M_{pl}$, $f_{env}$, and $F$.

This leaves specifying $p(M_{pl})$ and $p(f_{env})$.  First we address planet mass: while the result of Lop14 --- that sub-Neptune radii are relatively insensitive to their masses compared to the effect of the gaseous envelope mass fractions --- is what inspired this work, considering the planets' mass is still important for the smallest envelope fractions and thus for the posited transition between gaseous and rocky planets.  It is therefore necessary to retain consideration of the planet masses for this study.  Unfortunately, we do not have mass measurements for every individual planet in our complete subsample of \emph{Kepler}'s small planet candidates, and so we cannot specify a per-planet probability distribution for $M_{pl}$ as we did for $R_\star$.  However, we do have an idea of the mass distribution of the low-mass planet \emph{population} from radial velocity surveys.  Therefore, we can base our individual planet mass prior on the population distribution of masses; if we follow the RV surveys and choose a power law that is parameterized with the index $\alpha$, then:
\begin{align} \label{massprior}
p(M_{pl})& = \int p(M_{pl}|\alpha)p(\alpha) d\alpha \nonumber \\
 &= C \int M^\alpha p(\alpha) d\alpha .
\end{align} \\
Parameterizing the prior of an individual quantity based on the distribution within the population is exactly what makes this particular Bayesian formalism hierarchical, and is why we have turned to HBM to solve this problem.

\subsection{Hierarchical Bayesian Modeling}\label{HBM}

Mathematically, the general framework of HBM is a deceptively simple adjustment to Equation \ref{bayes}:

\begin{align}
p(\bm{\theta},\bm{\beta}|\bm{X}) = \frac{p(\bm{X}|\bm{\theta},\bm{\beta})p(\bm{\theta}|\bm{\beta})p(\bm{\beta})}{p(\bm{X})} ,\label{hiereqn}
\end{align} \\
where the difference between the set of individual parameters $\bm{\theta}$ and the set of population parameters $\bm{\beta}$, referred to as ``hyperparameters", has been made explicit.  Nevertheless, this rewrite, which is based only on the definition of conditional probability, makes a substantial difference in the interpretation of the problem, as one can now group individuals into populations, which both facilitates the characterization of the individual and allows the individual data to inform the population hyperparameters.  HBM thereby allows simultaneous inferences on both the parameters of the individual and of the population.

That said, HBM is not always necessary to answer the question that has been posed.  In particular, the hyperparameters may not always be of interest, in which case they can be treated as ``nuisance parameters" and marginalized over, as Equation \ref{massprior} implies.  Therefore, many hierarchical structures such as that in Figure \ref{basicstruct} do not necessarily need an HBM treatment.  The aspect of our problem which does require HBM is the specific question we have asked regarding compositions: because we want to infer the population distribution of compositions, we are interested in the analogous hyperparameters for $f_{env}$, and need the posterior to contain their distribution.  Only HBM can provide such a posterior that incorporates both the parameters of the individual planets and the population hyperparameters.

Applying this general framework specifically to $f_{env}$ means that we replace $p(f_{env})$ with $p(f_{env}|\mu,\sigma)p(\mu,\sigma)$, where $\mu$ and $\sigma$ are the hyperparameters characterizing the composition distribution.  The combination of computational convenience, the need for a distribution that can span several orders of magnitude, and the intuition that there should be fewer sub-Neptune planets with high envelope fractions leads us to choose a lognormal distribution for $f_{env}$, so that $\mu$ and $\sigma$ are the mean and standard deviation of the population of log($f_{env}$) values.  

This distribution does not factor in the expectation that significantly irradiated planets should have lost their envelopes, which we expect given the well-constrained rocky compositions of Corot-7b \citep{Jac10, Val10}, Kepler-10b \citep{Bat11,Kur13}, and Kepler-78b \citep{Pepe13,How13}.  Ignoring the physics of evaporation could therefore lead to an unrealistic composition distribution for these close-in planets.  That said, the photoevaporation of sub-Neptunes is an active area of theoretical research (e.g. \citealt{Owen12, Lop12, Lam13}), and so we err on the side of a simple yet theoretically motivated prescription to arrive at a realistic result.  In particular, we implement the mass loss threshold of \citet{Lop13}, which is a scaling law for the incident flux a planet would need to have received from its host star to have lost half of its initial H+He envelope after several Gyr ($F_{thresh}$); it is based on the assumption of energy-limited hydrodynamic escape and depends primarily on the mass of the planet's core, to a power that varies slightly depending on the planet's composition.  We model this irradiated sub-Neptune population by assigning a rocky composition ($f_{env}=0$) to a planet if 
\begin{align}\label{mlossthresh}
  F > F_{thresh} = (M_{core})^\gamma ,
\end{align}
where $F$ is the incident flux on the planet from its host star.  Otherwise, the planet has a non-zero $f_{env}$ which contributes to the composition distribution.  Given the theoretical uncertainties in this treatment of photoevaporation, we also allow $\gamma$ to vary, thereby adding a fourth hyperparameter to our model.

\subsection{Our Hierarchical Model}\label{ourHBM}

\begin{figure}[t]
\begin{center}
\begin{tikzpicture}[node distance = 1.5cm, auto]
    \node [param, fill=yellow!50] (mu) {$\mu$};
    \node [param, left of=mu, node distance = 2cm] (alpha) {$\alpha$};
    \node [param, right of=mu, node distance = 2cm, fill=yellow!50] (sigma) {$\sigma$};
    \node [plate,below of=mu, node distance = 3.1cm] (individtree) {
        \begin{tikzpicture}[anchor=center, node distance = 1.5cm, auto]
            \node [param, fill=yellow!50] (fenv) {$f_{env,i}$};
            \node [param, left of=fenv, node distance = 2cm] (mpl) {$M_{core,i}$};
            \node [param, right of=fenv, node distance = 2cm] (flux) {$F_i$};
            \node [param, below of=fenv] (rpl) {$R_{pl,i}$};
            \node [param, left of=rpl, node distance = 2cm] (rstar) {$R_{\star,i}$};
            \node [obsable, right of=rpl, node distance = 2cm] (deptherr) {$\sigma_{\delta,i}$};
            \node [obsable, below of = rpl] (depth) {$\delta_i$}; 
            \path [line] (fenv) -- (rpl);
            \path [line] (mpl) -- (rpl);
            \path [line] (flux) -- (rpl);
            \path [line] (rpl) -- (depth);
            \path [line] (rstar) -- (depth);
            \path [line] (deptherr) -- (depth);
        \end{tikzpicture}
    };
    \node [ghost, below of=alpha, node distance = 1.6cm] (mplg){};
    \node [ghost, below of=mu, node distance = 1.6cm] (fenvg){};
    \node [ghost, below of=sigma, node distance = 1.6cm] (fluxg){};
    \node [param, right of=fluxg, node distance = 2.4cm] (gamma) {$\gamma$};
    \node [ghost, below of=fenvg] (rplg){};
    \node [ghost, below of=sigma, text width=0.0cm, minimum height=0.0cm, node distance = 4.9cm, xshift=0.65cm] {$N$};
    \path [line] (alpha) -- (mplg);
    \path [line] (gamma) -- (rplg);
    \path [line] (mu) -- (fenvg);
    \path [line] (sigma) -- (fenvg);
\end{tikzpicture} \vspace{4mm}
\caption[blah]{The graphical representation of our final hierarchical model (see Equation \ref{statmod} for details).  The hyperparameters, i.e. those which define population-wide distributions, are located outside the rectangle (called a ``plate"), which represents the structure of individual parameters and data that is repeated for all of the planets in our sample ($i=1,...,N$=215).  The yellow parameters are of interest in this work; they are constrained by the observed data (gray) through MCMC simulations (Section \S \ref{JAGS}).\\
    \begin{itemize}[label={}]
       \item Data ($\bm{X}$):
       \begin{itemize}[label={}]
          \item $\delta_i$ = modeled transit depth
          \item $\sigma_{\delta,i}$ = transit depth uncertainty
       \end{itemize}
       \item Parameters ($\bm{\theta}$):
       \begin{itemize}[label={}]
          \item $M_{core,i}$ = mass of Earth-like rocky core
          \item $f_{env,i}$ = fraction of total mass in H+He envelope
          \item $F_i$ = incident stellar flux 
          \item $R_{pl,i}$ = planet radius
          \item $R_{\star,i}$ = stellar radius
       \end{itemize}
       \item Hyperparameters ($\bm{\beta}$):
       \begin{itemize}[label={}]
          \item $\alpha$ = index of the $M_{core,i}$ power law distribution
          \item $\mu$ = mean of the $f_{env,i}$ lognormal distribution
          \item $\sigma$ = standard deviation of the $f_{env,i}$ lognormal
          \item $\gamma$ = exponent of the envelope mass loss threshold
       \end{itemize}
    \end{itemize}
}  \vspace{-4mm} \label{fullstruct}
\end{center}
\end{figure}
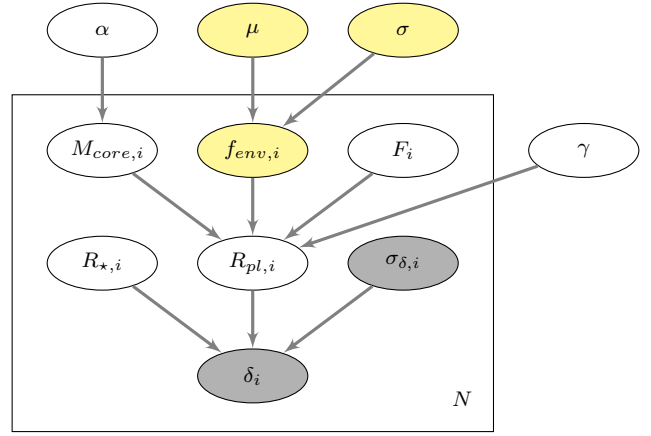

When written in the context of Bayes' Theorem, our full statistical model is the following: \\
\begin{align}  \label{fulleqn}
& p(\bm{\theta},\bm{\beta}|\bm{X}) \nonumber \\
& \qquad \propto \prod_{i=1}^N\Big\{p(\delta_i|\sigma_{\delta,i},R_{pl,i},R_{\star,i},M_{core,i},f_{env,i},F_i,\alpha,\mu,\sigma,\gamma)\Big\} \nonumber \\
& \qquad \qquad \times \prod_{i=1}^N\Big\{p(R_{\star,i})p(M_{pl,i}|\alpha)p(f_{env,i}|\mu,\sigma)\Big\} \nonumber \\
& \qquad \qquad \times p(\alpha)p(\mu)p(\sigma)p(\gamma) ,
\end{align} \\
where $\bm{X}=\{\delta_i,\sigma_{\delta,i}\}, \bm{\theta} = \{R_{pl,i},R_{\star,i},M_{core,i},f_{env,i},F_i\}$ and $\bm{\beta}=\{\alpha,\mu,\sigma,\gamma\}$ are defined in the caption of Figure \ref{fullstruct} and in the above text, and we are now considering all of the planet candidates, with sample size N=215.  Note that the normalizing constant $p(\bm{X})$ has not been written down, necessitating the expression of proportionality, and that we have assumed that all of the hyperparameters are independent of each other.  To illuminate how this follows Bayes' Theorem, we point out that the first line of this equation is the posterior, the second line is the likelihood, the third line contains the prior distributions for the individual parameters, and the fourth line contains the priors on the hyperparameters.

Needless to say, this equation is unwieldy, as is the case with most hierarchical models.  Graphical representations are therefore more often used to succinctly communicate the problem; the graphical model corresponding to Equation \ref{fulleqn} is shown in Figure \ref{fullstruct}.  However, neither equation nor figure contain the details of the various probability distributions, and so we introduce another, more informative way of writing down our hierarchical model.  In what follows, the quantities on the left-hand side are sampled from the distributions on the right-hand side; in other words, ``$q \sim$" is shorthand for ``$p(q) = $" where $p(q)$ is the probability distribution of the quantity q.  The parameters which directly specify each probability distribution are located after the ``$|$": \\
\begin{align} \label{statmod}
\delta_i|\sigma_{\delta,i},\bm{\theta},\bm{\beta} & \sim \text{Normal}\Big(\delta_i\Big|(R_{pl,i}/R_{\star,i})^2,\sigma_{\delta,i}^2\Big) \nonumber \\
R_{pl,i}|M_{core,i},f_{env,i},F_i,\bm{\beta} & = g(M_{core,i},f_{env,i},F_i,\gamma) \nonumber \\
R_{\star,i} & \sim \text{Gamma}\Big(R_{\star,i}\Big|a_i,b_i\Big) \nonumber \\ 
f_{env,i} | \mu, \sigma  & \sim \text{LogNormal}\Big(f_{env,i}\Big|\mu, \sigma \Big) \nonumber \\
M_{core,i} | \alpha & \sim \text{Pareto}\Big(M_{core,i}\Big| -(\alpha+1) , 0.5 \Big) \nonumber \\
\mu & \sim \text{Uniform}( -3.5 , -1 ) \nonumber    \\
log(\sigma^2) & \sim \text{Uniform}( -4 , 2 ) \nonumber \\
\gamma & \sim \text{Uniform}( 1 , 4 ) \nonumber \\
-(\alpha+1) & \sim \text{Beta}(-(\alpha+1) | 2 , 2 )  
\end{align}

Equation \ref{statmod} shows the details of our hierarchical Bayesian model, with the likelihood of the transit depth given the radius ratio and the transit depth measurement error in the first line; the interior structure models of Lop14 which map various planet properties to radius in the second line; an analytic fit to the marginal Hub14 likelihood of each host star's radius in the third line; the priors on the individual planet property parameters in lines 4 - 5; and the priors on the hyperparameters in lines 6 - 9.  

We have followed common practice and assumed a normal distribution for our likelihood, which is the equivalent assumption that one makes when using $\chi^2$.  Specifically this means that we have assumed that the measured transit depth $\delta_i$ is normally distributed around the ``true" transit depth equal to the planet-star radius ratio squared, with standard deviation set by the error on the transit depth.  The internal structure models are the power-law approximations given in Lop14; we did not use the full grid of models as the computational cost of interpolating a multi-dimensional grid was prohibitive within JAGS.  Although this imposes a factor of $\sim 2$ theoretical uncertainty in our inferred $f_{env}$ values, the width of the $f_{env}$ posteriors is still dominated by the substantial radius uncertainties, and we proceed with the more computationally efficient choice.  Due to similar concerns, we approximated the Hub14 stellar radius likelihoods that had been marginalized over all other stellar parameters as a gamma distribution by fitting its parameters $a$ and $b$ to each individual star.  

As discussed above, we have assumed a lognormal distribution for the planets' gaseous envelope mass fractions, where $\mu$ and $\sigma$ are the mean and standard deviation of the population of log($f_{env}$) values.  For the planet core masses we also follow convention and use a power-law distribution, which is known as a Pareto distribution in statistics; it is parameterized by the power-law index $\alpha$ and a lower limit which we have set to 0.5 M$_\oplus$.  We have truncated this power law so that all $M_{core,i} < 20$ M$_\oplus$; this is motivated by both the work of \citet{Mar14}, who find that planets in this size range have total masses between the mass detection threshold and 15-20 M$_\oplus$ (see Figure 49 of that paper), and the measurement of the most massive dense super-Earth found to date, Kepler-10c, at $M_{pl} \approx 17 \pm 2$ M$_\oplus$ \citep{Dum14}.

As for the priors on the hyperparameters, we use a uniform distribution for the ``location" parameter $\mu$ and a log-uniform distribution for the ``scale" parameter $\sigma^2$.  These distributions are equivalent to Jeffreys prior for these parameters and thus represent non-informative prior information (note that the uniform distribution is not always non-informative, especially for scale parameters or under parameter transformations).  For the core mass power law index $\alpha$, which must be $> -1$ for the power law to be proper\footnote{A proper probability distribution cannot integrate to $\infty$ over its support.}, we use previous results from radial velocity surveys (i.e. \citealt{How10}) and the intuition that smaller core masses must be more common to limit $0 < -(\alpha+1) < 1$ with diffuse but higher probability density around 0.5 (an index transformation is needed due to the way statisticians define the Pareto distribution).  This is naturally accomplished with the Beta distribution\footnote{The Beta distribution is defined as $p_{Beta} (x|\alpha_B,\beta_B) \propto x^{\alpha_B-1} (1-x)^{\beta_B-1}$ so that $p_{Beta} (x|2,2). \propto x (1-x)$}  whose parameters have both been set to 2.  Finally, we allow for theoretical uncertainties in the evaporation threshold power law index by allowing $\gamma$ to vary under a uniform prior distribution.

\subsection{JAGS: MCMC with Hierarchical Models}\label{JAGS}

Having fully specified this hierarchical model and motivated our choices for specific distributions, we can now run Markov Chain Monte Carlo (MCMC) simulations to give us posteriors on all of our parameters of interest.  Rather than write our own MCMC sampler, we use JAGS (Just Another Gibbs Sampler\footnote{JAGS code and user manuals can be downloaded at http://sourceforge.net/projects/mcmc-jags/.}; \citealt{Plu03}), which was written specifically to analyze hierarchical Bayesian models via MCMC.  Its platform independence and compatibility with the R computing language builds upon the BUGS project \citep{Lunn00}, which was historically developed for analyzing hierarchical models on Windows platforms.

As its name suggests, JAGS uses Gibbs sampling to proceed from step to step in the Markov chain, which requires the ability to write down the full conditional probability distribution of each parameter.  In practice, JAGS assigns different distributional families of Gibbs samplers to each parameter based on which sampling method is most efficient for the families of distributions involved in that region of the hierarchical model.  Many full conditionals are algebraically complicated and become expensive to evaluate, in which case JAGS implements Adaptive Rejection Metropolis Sampling.   The accepted parameter values are then stored and interpreted as samples from the target posterior distribution.

To produce the results shown in \S \ref{Res}, we run our model with 10 chains, each for 500,000 iterations.  The first half of each chain is discarded as ``burn-in", and the resulting half is thinned by a factor of 250, such that we retain 10,000 posteriors samples of each parameter.  JAGS computes the Gelman-Rubin convergence diagnostic \citep{Gel92} at run-time; the convergence of our MCMC simulations is analyzed in \S \ref{convshrink}.

\section{Results} \label{Res}

Here we present the results of the hierarchical MCMC simulations for the parameters of interest in this work (highlighted yellow in Figure \ref{fullstruct}).

\subsection{Population Composition Distribution} \label{compdist}

\begin{figure*}[t]
\begin{center}
\includegraphics[angle=270,scale=0.69]{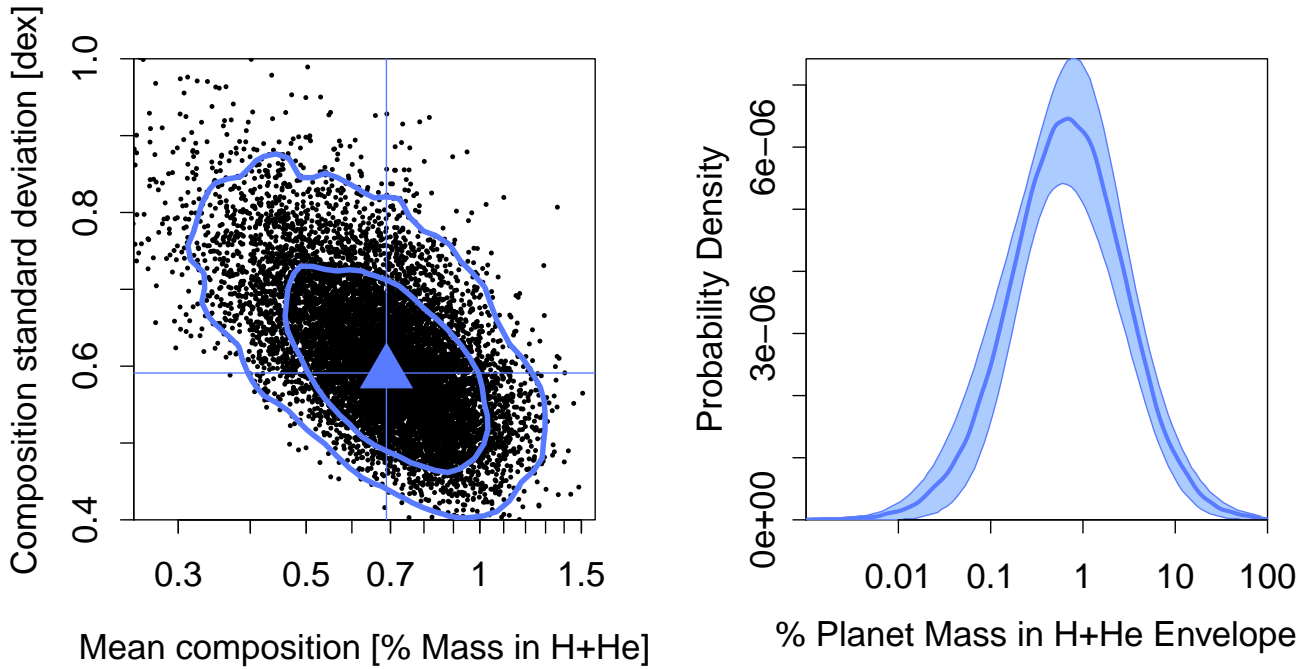}
\caption{\emph{Left:} Marginal posterior distribution for the mean $\mu$ and standard deviation $\sigma$ of the log($f_{env}$) population distribution, with 68\% and 95\% contours.  The ``best-fit" mean log envelope fraction, denoted by the vertical line at the large triangle, is -2.2 ($f_{env} \approx 0.007$).
\emph{Right:}  The posterior predictive composition distribution of \emph{Kepler}'s sub-Neptune planet candidates (solid line), with a 68\% coverage band.  The peak of this $f_{env}$ distribution corresponds to the ``best-fit" value of $\mu$, which shows that planets with 1 R$_\oplus < R_{pl} < 4$ R$_\oplus$ and an incident stellar flux low enough to retain a gaseous envelope are most likely to be composed of $\sim 1\%$ H+He by mass.} \label{fenvpost}
\end{center}
\end{figure*}

Figure \ref{fenvpost} shows the results for the top-most level of our model (see \S \ref{ourHBM}): the population-wide composition parameters.  More specifically, the left panel displays the marginal posterior distribution for $\mu$ and $\sigma$; these hyperparameters determine the mean and standard deviation, respectively, of the population distribution of log($f_{env}$) values.  As $f_{env}$ denotes the fraction of a planet's mass that exists in a hydrogen and helium envelope around an Earth-like rocky core, these parameters set the composition distribution of \emph{Kepler}'s sub-Neptune planet candidates under our assumed internal structure.  The ``best-fit" $\mu$ and $\sigma$ values, i.e. the mode of this posterior, are denoted by the large triangle and correspond to -2.2 dex ($\approx 0.7\%$) and 0.6 dex, respectively; they were found by performing two-dimensional kernel density estimation on a 50x50 grid and identifying the grid point with the highest density of posterior samples.   This 2D KDE also gives us the drawn contours enclosing 68\% and 95\% of the posterior density.  As the points in Figure \ref{fenvpost} range over the allowed values of every parameter, utilizing all of the posterior samples in the above calculation effectively marginalizes over all of the other parameters in Figure \ref{fullstruct}.

To elucidate what the hyperparameter posterior implies for the sub-Neptune composition distribution, we must map the allowed $(\mu, \sigma)$ values onto $f_{env}$ space.  This is shown in the right panel of Figure \ref{fenvpost}, where we plot the posterior predictive distribution of log($f_{env}$) in solid blue.  This distribution is computed by drawing (with replacement) 10,000 sets of ($\mu$,$\sigma$) values from the posterior in the left panel of Figure \ref{fenvpost}, which defines 10,000 $f_{env}$ distributions.   From each of these we randomly draw one log($f_{env}$) value; combining all of these values into one histogram effectively marginalizes over the uncertainty in $\mu$ and $\sigma$ and produces the log($f_{env}$) posterior predictive distribution.  To compute the 68\% coverage band in light blue\footnote{The coverage band is analogous to a confidence band in frequentist statistics, albeit with the requisite difference in interpretation, as it represents parameter uncertainty rather than variation between data sets.}, we draw several thousand log($f_{env}$) values from each set of ($\mu$,$\sigma$), which results in 10,000 log($f_{env}$) histograms.  On a bin-by-bin basis, we then find the range of counts which enclose 68\% of the histograms.  Note that this distribution does not include rocky planets; see Figures \ref{fenvrad}, \ref{rainbowhist}, and \ref{fracrocky} for discussion about the gas-rock transition.  Additionally, $f_{env} \sim 0.1\%$ corresponds to a gaseous envelope that extends $\sim 0.1$ R$_\oplus$ above the rocky core, which is below the radius precision for these planets; the constraints on the distribution for the smallest $f_{env}$ arise from the lognormal assumption.

The posterior predictive distribution (right panel of Figure \ref{fenvpost}) illustrates that the most likely composition for these sub-Neptune planet candidates is $\sim 1$\% H+He by mass.  This distribution represents the probability that, given the currently observed planet sample, the next observed planet in our considered size range (1-4 R$_\oplus$) will have a certain envelope fraction.  It therefore marginalizes over planetary radius, meaning that the shape of the observed radius distribution for this complete subsample of planet candidates is encoded in the shape of the envelope fraction distribution.  It is important to note that this distribution does not predict a planet's envelope fraction based on its measured radius; rather, it gives the distribution of envelope fractions over the entire population of sub-Neptunes.  To see how well radius maps to composition for individual planets, see Figures \ref{fenvrad} and \ref{rainbowhist}.

\subsection{Individual Planet Compositions} \label{individcomp}

While Figure \ref{fenvpost} gives the marginalized population distribution of compositions and so does not facilitate inferences that use knowledge of an individual's radius, our hierarchical model enabled us to compute individual composition posteriors for the 215 planet candidates in our complete \emph{Kepler} subsample.  These posteriors are summarized in Table \ref{individradcomp}.  Matching an arbitrary \emph{Kepler} planet's radius and radius uncertainty\footnote{Uncertainties are expressed in terms of the coverage interval which encloses the central 68\% of the posteriors (68\% C.I.).} to those given in Table \ref{individradcomp} will give, to first order, the range of allowed compositions for that planet.  Note that the radii given here are not exactly the same as the radii reported at the NExSci Exoplanet Archive, as the latter values do not use the full Hub14 stellar radius likelihood like we do here (also, see discussion about shrinkage in \S \ref{convshrink}).  Additionally, there will be some variation in composition for individual planets that are at different periods or that are hosted by stars of different spectral types, as these parameters do affect composition but to a lesser degree than radius.   We include periods and stellar radii in Table \ref{individradcomp} for these more detailed comparisons.

\begin{figure*}[t]
\begin{center}
\includegraphics[angle=270,scale=0.75]{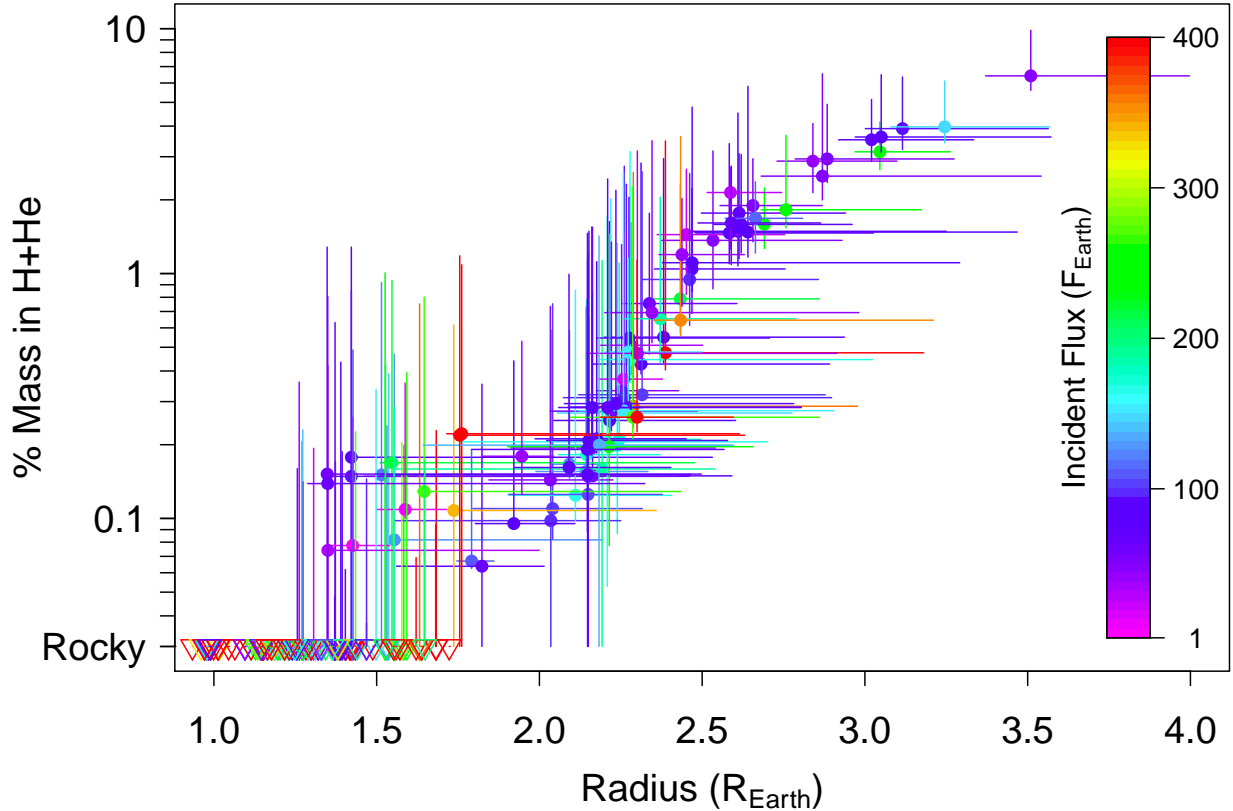}
\caption{Individual planet compositions as a function of radius.  Points denote the mode of the $f_{env}$ and $R_{pl}$ posteriors, while the lines denote the central 68\% coverage interval. Triangles denote rocky planets, for which more than half of the $f_{env}$ posterior samples are zero.  Color corresponds to the flux incident on the planet; we predict that the two planet candidates that have gaseous envelopes despite incident fluxes $F \ge 400$ F$_\oplus$ must have massive rocky cores, likely $> 10$ M$_\oplus$ (KOI 171.01 and KOI 355.01).  There is a locus of allowed compositions and radii, such that planets with $R_{pl} < 2$ R$_\oplus$ have $f_{env} < 1\%$, planets with $2< R_{pl} < 3$ R$_\oplus$ have $f_{env} \sim 1\%$, and planets with $R_{pl} > 3$ R$_\oplus$ have $f_{env} \sim$ a few \%.} \label{fenvrad}
\end{center}
\end{figure*}

Figure \ref{fenvrad} displays the information in Table \ref{individradcomp}, showing the individual planet compositions as a function of radius.   Points denote the mode of the $f_{env}$ and $R_{pl}$ posteriors, while the lines denote the central 68\% coverage interval (C.I.).  If more than half of an individual's $f_{env}$ posterior samples are zero, indicating that the stellar flux incident on the planet breached the mass loss threshold more than half of the time, we label that planet as ``rocky" and give it a triangular symbol.  For some of these planets, the 68\% C.I. includes non-rocky compositions; these planets have $f_{env}$ errors bars which extend into the gaseous region of parameter space.  Color corresponds to the flux incident on the planet, given the period and the stellar parameters reported in Hub14; red points denote planets with $F \ge 400$ F$_\oplus$.

\subsection{Posterior Checks and Convergence}\label{convshrink}

An important part of Bayesian analysis is testing for the convergence of the MCMC simulations, which we check for in a number of ways.  To begin, we compare the prior distributions for individual planet parameters to their posteriors for a quick yet illustrative reality check that our hierarchical MCMC simulations are producing reasonable results.  If we have strong prior information about the parameters, then the hyperparameter posteriors and the structure of the statistical model should preserve this information via posteriors that are similar in shape and location to the priors.  

\begin{figure}[t]
\includegraphics[angle=270,scale=0.4]{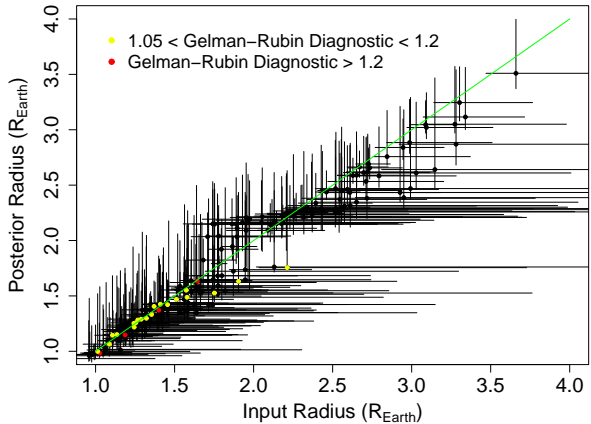}
\caption{Comparison of the ``prior" planet radius distributions to the posterior distributions, with convergence diagnostics for the corresponding $f_{env}$ posterior.  The $45^\circ$ line where posterior estimates equal prior estimates is green.  Each planet has a 68\% coverage interval that spans the one-to-one line, indicating that our model is behaving appropriately.} \label{radcheck}
\end{figure}

Figure \ref{radcheck} shows this check for the planet radii, which we treat as a parameter in our model and have strong prior information for (see Figure \ref{fullstruct} and Equation \ref{statmod}).  On the x-axis we plot the ``prior" radius distribution for each planet in our sample\footnote{The input planet radius ($R_{pl,i}$) distributions plotted here are not priors in the strictest sense of the definition, as the stellar radius is actually the quantity that has a distribution determined \emph{a priori}; however, since the transit depth uncertainties are small, the prior $R_{pl,i}$ distributions can be reasonably approximated as described here.}, which we compute by scaling the Hub 14 stellar radius likelihoods by the observed transit depth, and on the y-axis we plot the posterior radius distribution that result from our MCMC simulations.  The modes of the distributions are denoted as points, with the 68\% coverage interval spanned by the lines.  The color of the points denote the value of the Gelman-Rubin convergence diagnostic \citep{Gel92} for each planet's $f_{env}$ posterior, which we discuss in greater detail below.  The diagonal green line is the $45^\circ$ line, which we expect all of our individual radius distributions to span if our model is behaving as required.  We immediately see that this is the case, indicating that our model is incorporating our prior radius information appropriately and that our posteriors are accurate given our data and its assumed hierarchical structure.

We also see a few salient features of our model manifest in this figure.  First, there is a zone of avoidance between 1.7 and 2.0 R$_\oplus$ where the posterior radius distributions have been pushed to either side of the corresponding prior distributions (but not unreasonably so, given that each planet has a 68\% coverage interval spans the one-to-one line).  This is due to our incorporation of photoevaporation, as with periods $< 25$ days it is difficult to retain the tenuous gaseous envelope needed to produce these planetary radii.  Note that the planet radius priors are wide enough that this model feature does not pose significant problems for inferring a H+He envelope composition for each planet; however, if the planet parameters were more tightly constrained, as is the case for those orbiting brighter host stars with spectroscopic follow-up, a radius falling solidly within this narrow range would be better explained with a water-dominated composition, as we discuss in \S \ref{intromodels}.

Second, a general feature of hierarchical Bayesian models is evident in Figure \ref{radcheck}: the posteriors on these individual parameters are much narrower than the priors.  While this is expected for all Bayesian analyses given the role of data in constraining posterior estimates from prior information, the hierarchical structure of our problem contributes to this effect: by relating individuals to each other, HBM provides posterior estimates of individual exoplanet properties which have smaller variance than if multiple individual Bayesian analyses were performed independently.  This is called shrinkage, as this effect is achieved by ``shrinking" the posterior estimates toward the population mean (see \citealt{Lor07} for a more detailed discussion).  Indeed, Figure \ref{radcheck} shows that the points above 2.5 R$_\oplus$ fall slightly below the one-to-one line, and the points below 2 R$_\oplus$ fall slightly above it, as the mean envelope fraction of about 1\% roughly corresponds to a radius of $\sim 2.2$ R$_\oplus$.

We continue the discussion of convergence with Gelman-Rubin convergence statistic ($\hat{R}$) for the individual $f_{env}$ posteriors.  This diagnostic calculates the ratio of the total variance across all chains in the MCMC simulation to the variance within individual chains; $\hat{R}$ within a percent or so of 1 indicates convergence, where each individual chain probes about the same volume of parameter space as all of the chains taken together.  Most of the individual $f_{env}$ posteriors have $\hat{R} \leq 1.01$, but there are some planets whose $\hat{R}$ values indicate that the MCMC should be run longer.  One immediately notices that these planets have small radii; in fact, every planet with $\hat{R} > 1.01$ has a $f_{env}$ posterior that spans 0.  Given the discrete nature of the switch between rocky and gaseous compositions, the fact that mixing between $f_{env}$ chains is worse for the planets which cross this transition is not a surprise; additionally, we expect $\hat{R}$ to be biased high simply as a numerical artifact of representing rocky compositions with $f_{env} = 0$, as this imposes a gap between rocky and non-rocky $f_{env}$ chains.  Noting that the planets with rocky compositions do not contribute to constraints on the composition distribution hyperparameters, we conclude that these $\hat{R}$ values are not a cause for concern.

\begin{figure*}[t]
\begin{center}
\includegraphics[angle=270,scale=0.7]{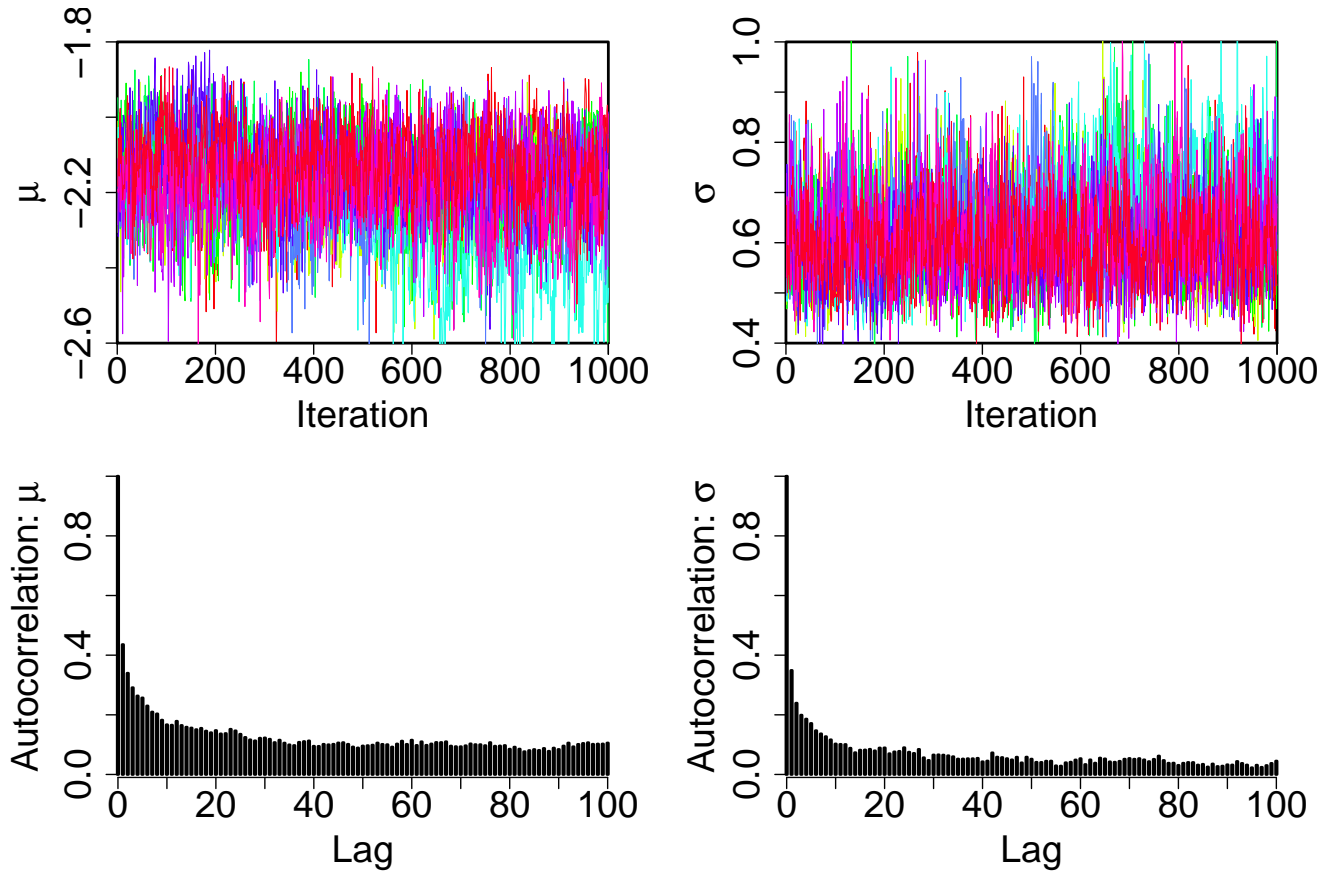}
\caption{Graphical convergence diagnostics for our composition hyperparameters $\mu$ (left column) and $\sigma$ (right column); the first row displays the value of all 10 MCMC chains as a function of location in the chain (a ``trace plot"), while the bottom row displays the average chain autocorrelation at increasing offsets.  Both parameters have good mixing between the chains and quickly reach a low level of autocorrelation, indicating that we have converged to the stationary distribution for the joint posterior displayed in Figure \ref{fenvpost}.} \label{hypertrace}
\end{center}
\end{figure*}

Similarly, we must assess the convergence of the composition hyperparameters in our simulation.  The $\hat{R}$ values for $\mu$ and $\sigma$ are 1.08 and 1.03, respectively.  However, as we see above, $\hat{R}$ does not always convey the full picture of convergence, so we turn to other common diagnostics such as trace plots and autocorrelation functions (Figure \ref{hypertrace}, top and bottom rows respectively) to more fully investigate the issue.  In the trace plots, the values of a parameter's chain is plotted as a function of location along the chain, with different colors indicating different chains.  We see that there is good mixing between the chains for both parameters, indicating that we have arrived at the stationary distribution for the joint posterior displayed in Figure \ref{fenvpost}.  

The average chain autocorrelation functions further support the convergence of our simulations, as they quickly reach a low level of autocorrelation.  The slightly higher autocorrelation present in $\mu$ explains the slightly higher $\hat{R}$ calculated for that parameter, but the difference is not strong enough to be visible in the trace plots.  Given that the mode of the $\mu$ posterior distribution has been well established through the mixing of the existing chains, our conclusion that the most likely sub-Neptune composition is $\sim 1\%$ H+He by mass would not change by running the simulation longer.  With such diminishing returns regarding convergence, we take our ($\mu$,$\sigma$) posterior as the stationary distribution and continue with a discussion of these results.

\section{Discussion} \label{Discuss}

The results in \S \ref{Res} have numerous implications for characterizing the sizable sub-Neptune population discovered by \emph{Kepler}.  We discuss several in detail below, but first we address some of the more constraining choices that we have made in our statistical model (see \S \ref{basicBayes} - \ref{ourHBM} for further description and motivation for all of the assumptions we make).

\subsection{Motivation for Salient Model Assumptions}

Arguably the most obvious assumption we've made is that the composition distribution can be reasonably described with a lognormal distribution.  There is currently very little theoretical guidance regarding the expected shape of this distribution; in the absence of such predictions, we turn to our driving science questions (see \S \ref{intro}) to inform this choice.  Because we are interested in characterizing the ``typical" sub-Neptune envelope fraction as well as the range present in the population, the most natural choice is a distribution that straightforwardly parameterizes the mean and variance of a population: a normal distribution.  In addition, we expected a large dynamic range of gaseous envelope fractions, which the lognormal in particular is able to accommodate.  We acknowledge that different choices for this composition distribution can affect the result we present here, as the particular parametric form drives the quantitative details of the shrinkage we observe in \S \ref{convshrink}.  Alternatively, one could completely sidestep this concern by adopting a nonparametric approach; however, doing so involves solving for a much larger number of free parameters, which simultaneously reduces the predictive power and expands the computational expense of such a study.  Our choice therefore best balances the demands of our scientific goals, our computational considerations, and the desire to limit the number of free parameters in an already fairly complex statistical model.

We also make several assumptions that are not explicit in our statistical model.  First, we assume that all of the planet candidates in our sample are true planets.  If we were concerned with an absolute occurrence rate of planet compositions, we would need to correctly account for the presence of false positives; however, in this work we are interested in the shape and location of the composition distribution and can safely ignore the normalization constant needed for occurrence rate studies.  For our purposes it is therefore sufficient to note that the probability of a given planet candidate being a false positive is roughly constant over our radius range ($\sim 5-10\%$; see \S \ref{incomplete}), and so the presence of false positives are not expected to affect our results.  

\begin{figure*}[t]
\begin{center}
\includegraphics[angle=270,scale=0.7]{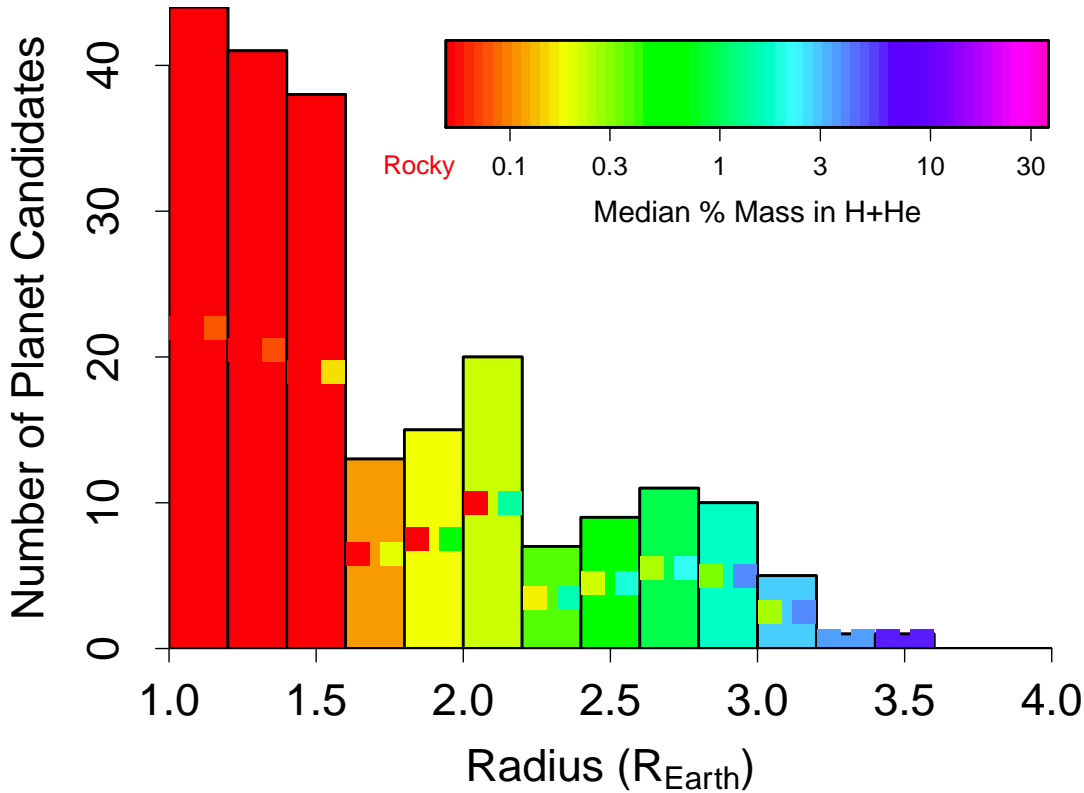}
\caption{The radius distribution of our complete subsample, color coded according to the median composition in each bin.  The squares denote the full range of compositions within the bin: the lowest $f_{env}$ in each bin corresponds to the left colored box, and the highest $f_{env}$ the right.  On average, interpreting radius as a proxy for composition is reasonable, although the large radius errors do allow some dispersion.} \label{rainbowhist}
\end{center}
\end{figure*}

Second, we do not correct for transit probability, which is acceptable if planet composition is uncorrelated with period.  There are a number of reasons to expect that this correlation could exist due to the conditions under which these planets form and evolve; even our own incorporation of photoevaporation predicts a slight dependence between incident flux and composition (see Figure \ref{fracrocky}).  This is a very interesting area for future work but is outside the scope of this study as it requires modeling the underlying period distribution of these planets, which would have added an additional layer of complexity to Figure \ref{fullstruct}.  In this work, our prescription for photoevaporation does introduce some period-flux dependence, but fortuitously  our results remain insensitive to the transit probability correction due to the lack of correlation between $\gamma$, which controls the rock-gas flux transition (Equation \ref{mlossthresh}), and the composition hyperparameters.  Note that we incorporate $\gamma$, which varies between 2.2 and 3.0 in our posterior samples, as a free parameter to account for theoretical uncertainty in this threshold, so much of the period-composition dependence present for individual planets gets washed out over the marginalized $f_{env}$ posteriors presented in Table \ref{individradcomp}.

Finally, we revisit the possibility that some of these planets may be water worlds.  In \S \ref{intromodels} we motivated from an observational perspective why we assume all of these planets have a composition consisting of a rocky core and a hydrogen/helium envelope: for the first investigation of this population's composition distribution, it is natural to extend a two-component composition to the mid-range planetary sizes between the small, highly irradiated planets known to be rocky and the large low-mass sub-Neptunes must have at least some H+He.  Given these limits, it is difficult to motivate a population of sub-Neptunes that must all be characterized as water worlds.  However, this does not mean that there cannot be a sub-population of water worlds, especially at periods longer than the planet candidates we consider in this work, and so this proposal is rich in possibilities for future work.  Nevertheless, given the degeneracy (discussed in \S \ref{intromodels}) between detailed compositions and measured mass and radius, an additional observable that can reliably distinguish between water-poor and water-rich bulk compositions will need to be measured and introduced to a statistical model like this one in order to get a quantitative handle on the extent of this possible sub-population.

\subsection{Radius as a Proxy for Composition} \label{radproxy}

A locus through ($R_{pl},f_{env}$) space is immediately apparent in Figure \ref{fenvrad}, illustrating that ``radius as a proxy for composition" is a reasonable interpretation to adopt for planets with $R_{pl} > 2$ R$_\oplus$, even with the current large, asymmetric errors on the planet radii.  However, more variability is evident for smaller planets, especially in the $1.2 < R_{pl} < 1.8$ R$_\oplus$ range, where planets can either have rocky or gaseous compositions (see \S \ref{rockgastrans} for a more detailed discussion).  Given that the realistic radius errors included in this study does widen this locus, the following summary provides a reasonable rule-of-thumb when interpreting the composition of planets based on their radii: planets with $R_{pl} < 2$ R$_\oplus$ have $f_{env} < 1\%$, planets with $2< R_{pl} < 3$ R$_\oplus$ have $f_{env} \sim 1\%$, and planets with $R_{pl} > 3$ R$_\oplus$ have $f_{env} \sim$ a few \%.

Figure \ref{rainbowhist} further illustrates how the strong monotonic relationship between radius and composition can be extended to interpreting compositions from an observed radius histogram.  We plot the radius distribution of our complete subsample (also shown in blue in Figure \ref{samp}), but now color-code each bin according to the median composition of those planets, where a single value for composition, the mode of the $f_{env}$ posterior, has been used for each planet.  Taking radius as a proxy for composition would result in a monotonic increase in composition across the bins, which is exactly what we see.  The picture complicates a bit when we consider the full range of compositions present in each bin, as illustrated by the colored boxes: the color of the left box corresponds to the lowest $f_{env}$ in that bin, and the color of the right box corresponds to the highest $f_{env}$.  The range within each bin illustrates the dispersion accommodated by the substantial errors on the planet radii.  While the dispersion is currently non-negligible, it does not disrupt the average relationship between radius and composition.

\subsection{The Rock-Gas Transition} \label{rockgastrans}

Figure \ref{fenvrad} also has implications for the expected transition between rocky and gaseous planets, assuming these planets do not have an appreciable mass fraction of water.  In particular, we see that planets with $1.2 < R_{pl} < 1.8$ R$_\oplus$ can be either rocky or gaseous, with $f_{env}$ posteriors that span both compositions.  This is consistent with the finding of \citet{Rog14}, which places the transition between rocky and gaseous planets at 1.5 R$_\oplus$ based on $\sim 50$ \emph{Kepler} confirmed planets with radial velocity mass measurements, primarily from \citet{Mar14}.  It is notable that internal structure models combined with the back-of-the-envelope parametrization of photoevaporation that we employ here (Equation \ref{mlossthresh}) is able to reproduce this result within the context of these hierarchical MCMC simulations, given that we use no mass measurements to provide constraints as does \citet{Rog14}.

\begin{figure}[t]
\includegraphics[angle=270,scale=0.43]{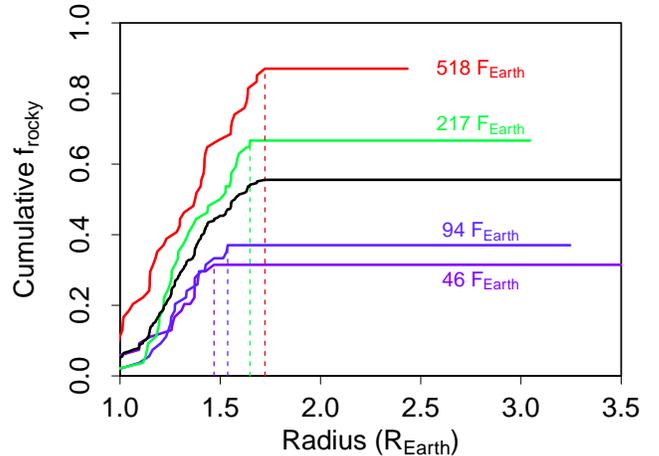}
\caption{The cumulative fraction of planets that are rocky in four flux bins colored at the same scale as Figure \ref{fenvrad} according to their labeled median flux values.  Each colored bin contains 54 planets; the black line is the cumulative fraction for the entire sample.  A planet is considered rocky if more than half of its $f_{env}$ posterior occurs at 0.  We see that the maximum radius for a rocky planet, denoted by the dotted vertical lines, increases slightly with increasing incident flux.} \label{fracrocky}
\end{figure}

Our implementation of photoevaporation further predicts that there is some flux dependence to this transition, as seen in the color variation as a function of radius for the planets that could have either composition.  Figure \ref{fracrocky} more clearly illustrates this dependence: we plot the cumulative fraction of planets that are rocky in four flux bins, each containing 54 planets; the black line is the cumulative fraction for the entire sample.  A planet is considered rocky if more than half of its $f_{env}$ posterior occurs at 0, as is the case for the triangles in Figure \ref{fenvrad}.  We see that the maximum radius for a rocky planet, denoted by the dotted vertical lines, increases slightly with increasing incident flux.

\citet{Rog14} addressed this possibility by computing the marginal likelihood of the data under different hypothetical gas/rock transitions, including a sharp step function, a gradual linear relationship for the fraction of rocky planets as a function of radius, and a transition that depended on incident flux.  With the existing large mass uncertainties, they find that the sharp transition is slightly favored over both other options with a Bayes factor of $\sim 2$.   We note that this Bayes factor is actually quite small for the purposes of inference, as one's prior belief in the realism of each of these transitions can still be a large factor in inferring which model best reflects what happens in nature.  Furthermore, this factor can depend strongly on the choice of hyperprior, particularly when the prior is formally an improper distribution like the uniform distributions that were used.  Therefore, we echo the author's caution that this result does not mean the arguably more realistic transitions are ruled out, but rather that the currently large mass uncertainties do not allow one to distinguish between these possibilities.

Given the result of \citet{Rog14}, more precise mass measurements are needed before we can conclusively test the prediction that large rocky planets must have high incident stellar fluxes.  In particular, radial velocity follow-up of \emph{Kepler} planet candidates can most effectively contribute to our understanding of photoevaporation by targeting $1.2 < R_{pl} < 1.8$ R$_\oplus$ planets at incident fluxes near this flux threshold.  Two such planets are immediately identifiable in Figure \ref{fenvrad}, due to their high incident fluxes compared to the other similarly sized planets: KOI 171.01 (Kepler-116 b) and KOI 355.01, at 2.4 and 2.3 R$_\oplus$, and 470 and 440 F$_\oplus$, respectively.  Because the mass loss flux threshold, and therefore the retention of the planet's envelope, is dependent on the core mass of the planet, we predict these planets must have fairly massive rocky cores, likely $> 10$ M$_\oplus$.  These planet candidates also happen to have fairly bright host stars, at a \emph{Kepler} magnitude of 13.7 and 13.2, respectively, and so this prediction could be tested with radial velocity measurements.  KOIs 171.01 and 355.01 therefore provide excellent leverage for testing theories of photoevaporation.

Regarding planets with massive cores, it is interesting to note that the most massive dense super-Earth found to date, Kepler-10c \citep{Dum14}, would not in fact be rocky according to the models we use here.  Based on its measured mass and radius ($\approx 17 \pm 2$ M$_\oplus$ and 2.35 R$_\oplus$), Kepler-10c should have a gaseous envelope fraction of $\sim 0.5\%$ (Lop14), or a relatively massive water steam envelope.  Rather than representing an extreme on the spectrum of possible super-Earth compositions, Kepler-10c instead exemplifies what we predict to be a fairly typical if somewhat massive sub-Neptune in terms of the envelope mass fraction it could possess.

\subsection{No Population-Wide Mass-Radius Relationship} \label{noMRR}

\begin{figure*}[t]
\begin{center}
\includegraphics[angle=270,scale=0.75]{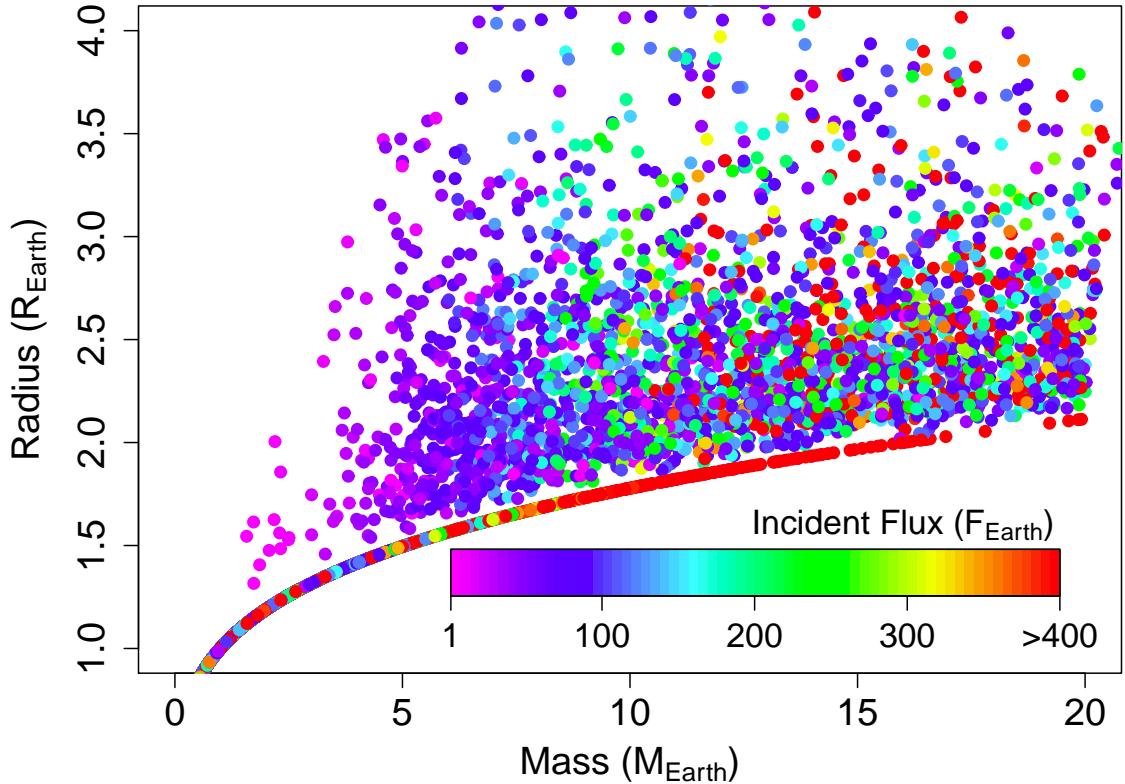}
\caption{The masses and radii of a population of 10,000 planets generated from our ``best fit" composition distribution (see Figure \ref{fenvpost}).  Each point is colored according to its incident flux, at the same scale as Figure \ref{fenvrad}.  Immediately we see that there is no clear mass-radius relationship for these sub-Neptune planets, although there are disallowed regions due to the maximum density of a rocky planet (at high masses and small radii) and to photoevaporation (at low masses and large radii).  There is also a higher number density of planets between 2 and 2.5 R$_\oplus$; this is a direct result of our composition distribution peaking around $f_{env} \sim 1\%$.} \label{MRgenplan}
\end{center}
\end{figure*}

Figure \ref{MRgenplan} illustrates what the sub-Neptune composition distribution that we find implies for the mass-radius relationship of these planets.   Specifically, we generate a population of 10,000 planets using our ``best fit" composition distribution ($\mu = 0.7\%; \sigma = 0.6$ dex) and a core mass distribution $\propto M^{-1}$, then randomly match these planets to the host stars and periods of the planets in our sample to apply the rock-gas transition flux threshold.  The color corresponds to the generated planets' incident flux as in Figure \ref{fenvrad}, with the gradation at the low-mass end arising from our core mass-dependent prescription for photoevaporation.  There is also a higher number density of planets between 2 and 2.5 R$_\oplus$; this is a direct result of our composition distribution peaking around $f_{env} \sim 1\%$.  

Immediately we see that there is no clear one-to-one relationship, although there is a disallowed region at high masses and small radii, which is due to the maximum density of a rocky planet, and another at low masses and large radii, which is due to photoevaporation.  While one could certainly fit a line to these points in mass and radius space, we argue that the more physically interesting variable at play is the composition.  Because having a range of compositions dominates the spread in this plot (that is, the vertical extent of the mass-radius ``relationship" is controlled by the distribution of compositions; see Figure \ref{rainbowhist} for another illustration of this), understanding planetary compositions in a population-wide sense requires robust statistical modeling that incorporates distributions rather than just mean relationships.

The lack of a mass-radius relationship for these sub-Neptune planets, which compose the majority of the planets that \emph{Kepler} has detected, also has major implications for dynamical studies which require \emph{Kepler} radii to be mapped to masses.  Namely, such studies must adopt a probabilistic approach to allow for a distribution of masses at a given radius.  Without a way to incorporate the dispersion between mass and radius, the authors could be mislead by results that are seemingly more precise than they actually are.  Similarly, theoretical studies could mistakenly rule out different parts of parameter space that may actually be allowed given the intrinsic uncertainty in the planet's mass based only on its radius.

\subsection{Implications for Population Formation Models} \label{popsynth}

Via our physically informed statistical modeling (\S \ref{ourHBM}) we have inferred the mean and variance of the present-day compositions of planets with 1 R$_\oplus < R_{pl} < 4$ R$_\oplus$, finding an average $f_{env}$ of $\sim 1\%$ and standard deviation of $\sim 0.5$ dex, respectively (\S \ref{compdist}).  As this result is derived directly from \emph{Kepler} data, it offers a strong observational constraint for studies of planet formation which strive to characterize not only the average behavior of a few planets, but the range and distribution of various physically interesting variables across an entire planet population.  Key diagnostics such as the range of compositions for these small planets can, for example, inform the degree of gas accretion during the planet formation process, and can therefore provide constraints on the relevant local protoplanetary disk parameters such as temperature and viscosity.  

Of course, planetary evolution could also affect these planets' present-day compositions, and so the composition distribution we infer here has also encoded information about any of these processes which may have occurred.  These quantitative constraints provide a first step in enabling comparisons between the effect that disk migration vs. multi-body interactions vs. in-situ formation could have on the amount of gas retained by super-Earths and sub-Neptunes, the most common kind of planets in our Galaxy.  Much work remains to be done to disentangle these effects, and many other observational indicators such as spin-orbit misalignment and period ratios within multi-planet systems are being scrutinized.  Nevertheless, with this analysis, planetary compositions can also enter into the conversation in a quantitative way.

\section{Conclusions} \label{Conclu}

In this paper we present the first quantitative distribution of sub-Neptune compositions.  We find that, if these planets are composed of an Earth-like rocky core with a hydrogen and helium envelope, the ``typical" sub-Neptune has $\sim 1$\% of its mass in the gaseous envelope, while the population has a spread of $\pm 0.5$ dex.  We arrive at this result by carefully choosing a subsample of \emph{Kepler} planet candidates (\S \ref{Kepdata}) that is complete above 1.2 R$_\oplus$ (\S \ref{sampsel}) and adopting a hierarchical Bayesian framework (\S \ref{method}) with a realistic yet relatively simple statistical model (\S \ref{ourHBM}) which incorporates the internal structure models of \citet{Lop14} and the stellar radius likelihoods derived by \citet{Hub14}.  This approach simultaneously accounts for the lack of mass measurements and substantial radius measurement errors while describing the population-wide behavior with only four free parameters.  

Our hierarchical Markov Chain Monte Carlo simulations (\S \ref{JAGS}) result in posteriors on both the compositions of individual planets (\S \ref{individcomp}) and on the composition distribution of the population (\S \ref{compdist}).  Therefore, in addition to finding that the mean and standard deviation of the present-day compositions of planets with 1 R$_\oplus < R_{pl} < 4$ R$_\oplus$ is $\sim 1\%$ and $\sim 0.5$ dex, we can identify an honest the rule-of-thumb that relates radius to composition: planets with $R_{pl} < 2$ R$_\oplus$ have $f_{env} < 1\%$, planets with $2< R_{pl} < 3$ R$_\oplus$ have $f_{env} \sim 1\%$, and planets with $R_{pl} > 3$ R$_\oplus$ have $f_{env} \sim$ a few \%.

Finally, we discuss the implications that these results have for various issues related to the compositions of sub-Neptune planets.  First, we verify that taking radius as a proxy for composition does hold up in the average sense even considering the large radius errors that exist for the majority of \emph{Kepler} planet candidates (\S \ref{radproxy}).  We also address the rock-gas transition and discuss how carefully chosen and precise mass measurements could help test the theory of photoevaporation by elucidating a transition that is a function of incident flux (\S \ref{rockgastrans}).  In \S \ref{noMRR} we illustrate how this composition distribution means that there is no mass-to-radius relationship for sub-Neptunes, and so dynamical studies must derive masses from \emph{Kepler} radii probabilistically rather than deterministically.  Finally, we discuss the rich opportunity these results offer for comparisons of planet formation studies with \emph{Kepler}'s observed planetary candidates.

\acknowledgments

We would like to thank the Statistical and Applied Mathematical Sciences Institute (SAMSI) for organizing a 3-week workshop on the Statistical Analysis of Kepler data, during which the primary author learned how to apply HBM and relevant computational techniques to research problems similar to this one, under the guidance of Merlise Clyde, Eric Ford, Jessi Cisewski, Robert Wolpert, and others.  Additionally, she would like to thank the members of the Bayesian Characterization of Exoplanet Populations group which arose out of this workshop, for their invaluable advice and feedback, particularly Leslie Rogers, Tom Loredo, Darin Ragozzine, Megan Shabram, Daniel Foreman-Mackey, and David Hogg, in addition to those listed above. Finally, we would like to thank Greg Laughlin and Jonathan Fortney for their advice and support as our graduate research advisors in the execution of this project.

The primary author's financial support during this investigation was primarily provided by the National Science Foundation Graduate Research Fellowship under Grant No. 0809125, with additional funding by NASA contract NNG14FC03C through subaward agreement 5710003702.  The secondary author would like to acknowledge support from NSF grant AST-1010017 and the UCSC Chancellor's Dissertation Year Fellowship.  This material was based upon work partially supported by the National Science Foundation under Grant DMS-1127914 to the Statistical and Applied Mathematical Sciences Institute. Any opinions, findings, and conclusions or recommendations expressed in this material are those of the author(s) and do not necessarily reflect the views of the National Science Foundation.  This research has made use of the NASA Exoplanet Archive, which is operated by the California Institute of Technology, under contract with the National Aeronautics and Space Administration under the Exoplanet Exploration Program.  This paper includes data collected by the Kepler mission. Funding for the Kepler mission is provided by the NASA Science Mission directorate.

{\it Facilities:} \facility{Kepler}.

\clearpage


\LongTables
\begin{deluxetable*}{cccccccc}
\tabletypesize{\footnotesize}
\tablecolumns{8}
\tablewidth{0pt}
\tablecaption{Compositions of Individual Planets in Sample \label{individradcomp}}
\tablehead{
\colhead{KOI \#} & \colhead{Kepler ID} & \colhead{$R_{pl}$}  & \colhead{68\% C.I.}  & \colhead{$P$}  & \colhead{$R_\star$} &  \colhead{$f_{env}$} & \colhead{68\% C.I.} \\
 & & \colhead{(R$_\oplus$)} & \colhead{(R$_\oplus$)} & \colhead{(days)} & \colhead{(R$_\odot$)} & \colhead{(\%)} & \colhead{(\%)}
}
\startdata
49.01 & 9527334 & 3.25 & ( 3.08 , 3.57 ) & 8.31 & 1.03 & 4.0 & ( 3.4 , 6.1 ) \\
69.01 & 3544595 & 1.63 & ( 1.59 , 1.69 ) & 4.73 & 0.92 & Rocky & ( 0 , 0 ) \\
70.01 & 6850504 & 3.05 & ( 2.97 , 3.57 ) & 10.85 & 0.92 & 3.6 & ( 3.1 , 6.5 ) \\
70.02 & 6850504 & 1.74 & ( 1.75 , 2.36 ) & 3.70 & 0.93 & 0.1 & ( 0.0 , 0.6 ) \\
70.05 & 6850504 & 0.98 & ( 0.92 , 1.18 ) & 19.58 & 0.91 & Rocky & ( 0 , 0 ) \\
82.01 & 10187017 & 2.59 & ( 2.52 , 2.74 ) & 16.15 & 0.76 & 2.1 & ( 1.2 , 2.7 ) \\
82.02 & 10187017 & 1.37 & ( 1.34 , 1.43 ) & 10.31 & 0.77 & Rocky & ( 0 , 0 ) \\
84.01 & 2571238 & 2.47 & ( 2.35 , 2.76 ) & 9.29 & 0.86 & 1.0 & ( 0.7 , 2.2 ) \\
103.01 & 2444412 & 2.88 & ( 2.79 , 3.27 ) & 14.91 & 0.91 & 2.9 & ( 2.3 , 4.9 ) \\
104.01 & 10318874 & 2.66 & ( 2.57 , 2.81 ) & 2.51 & 0.64 & 1.7 & ( 1.2 , 2.3 ) \\
112.02 & 10984090 & 1.30 & ( 1.22 , 1.59 ) & 3.71 & 1.06 & Rocky & ( 0 , 0 ) \\
116.01 & 8395660 & 2.46 & ( 2.32 , 2.86 ) & 13.57 & 1.02 & 0.9 & ( 0.6 , 2.5 ) \\
116.04 & 8395660 & 1.15 & ( 1.06 , 1.34 ) & 23.98 & 0.99 & Rocky & ( 0 , 0 ) \\
139.02 & 8559644 & 1.57 & ( 1.45 , 1.82 ) & 3.34 & 1.12 & Rocky & ( 0 , 0 ) \\
148.01 & 5735762 & 2.15 & ( 2.06 , 2.37 ) & 4.78 & 0.90 & 0.2 & ( 0.1 , 0.8 ) \\
148.02 & 5735762 & 3.02 & ( 2.92 , 3.33 ) & 9.67 & 0.88 & 3.5 & ( 2.8 , 5.1 ) \\
150.01 & 7626506 & 2.28 & ( 2.19 , 3.02 ) & 8.41 & 0.85 & 0.4 & ( 0.3 , 3.1 ) \\
153.01 & 12252424 & 2.44 & ( 2.37 , 2.63 ) & 8.93 & 0.71 & 1.2 & ( 0.7 , 2.0 ) \\
153.02 & 12252424 & 2.16 & ( 2.09 , 2.32 ) & 4.75 & 0.71 & 0.3 & ( 0.2 , 0.8 ) \\
157.01 & 6541920 & 3.12 & ( 3.00 , 3.56 ) & 13.02 & 1.09 & 3.9 & ( 3.2 , 6.3 ) \\
157.02 & 6541920 & 3.51 & ( 3.37 , 4.00 ) & 22.69 & 1.09 & 6.4 & ( 5.6 , 9.8 ) \\
157.06 & 6541920 & 2.14 & ( 1.99 , 2.33 ) & 10.30 & 1.07 & 0.2 & ( 0.1 , 0.7 ) \\
159.01 & 8972058 & 2.37 & ( 2.26 , 2.79 ) & 8.99 & 1.05 & 0.7 & ( 0.4 , 2.0 ) \\
159.02 & 8972058 & 0.99 & ( 0.93 , 1.24 ) & 2.40 & 1.04 & Rocky & ( 0 , 0 ) \\
161.01 & 5084942 & 2.69 & ( 2.63 , 2.84 ) & 3.11 & 0.82 & 1.6 & ( 1.2 , 2.2 ) \\
162.01 & 8107380 & 2.64 & ( 2.48 , 3.47 ) & 14.01 & 1.08 & 1.5 & ( 1.1 , 5.8 ) \\
165.01 & 9527915 & 2.66 & ( 2.56 , 2.87 ) & 13.22 & 0.81 & 1.9 & ( 1.3 , 2.9 ) \\
166.01 & 2441495 & 2.25 & ( 2.18 , 2.43 ) & 12.49 & 0.77 & 0.3 & ( 0.3 , 1.3 ) \\
167.01 & 11666881 & 1.76 & ( 1.74 , 2.63 ) & 4.92 & 1.19 & 0.2 & ( 0.0 , 1.2 ) \\
171.01 & 7831264 & 2.39 & ( 2.29 , 3.18 ) & 5.97 & 1.18 & 0.5 & ( 0.4 , 3.4 ) \\
171.02 & 7831264 & 2.21 & ( 1.76 , 2.70 ) & 13.07 & 1.16 & 0.2 & ( 0.0 , 1.7 ) \\
172.01 & 8692861 & 2.28 & ( 2.18 , 2.71 ) & 13.72 & 0.90 & 0.5 & ( 0.3 , 2.0 ) \\
177.01 & 6803202 & 2.22 & ( 2.03 , 2.49 ) & 21.06 & 1.14 & 0.3 & ( 0.2 , 1.3 ) \\
180.01 & 9573539 & 2.62 & ( 2.54 , 2.96 ) & 10.05 & 0.94 & 1.6 & ( 1.2 , 3.0 ) \\
238.01 & 7219825 & 2.58 & ( 2.48 , 3.03 ) & 17.23 & 1.11 & 1.5 & ( 1.1 , 3.4 ) \\
273.01 & 3102384 & 2.09 & ( 2.04 , 2.21 ) & 10.57 & 1.08 & 0.2 & ( 0.1 , 0.6 ) \\
280.01 & 4141376 & 2.15 & ( 2.09 , 2.26 ) & 11.87 & 1.04 & 0.2 & ( 0.1 , 0.6 ) \\
282.02 & 5088536 & 1.11 & ( 1.08 , 1.19 ) & 8.46 & 1.13 & Rocky & ( 0 , 0 ) \\
283.01 & 5695396 & 2.34 & ( 2.25 , 2.61 ) & 16.09 & 1.03 & 0.8 & ( 0.4 , 1.7 ) \\
299.01 & 2692377 & 1.42 & ( 1.36 , 1.59 ) & 1.54 & 0.94 & Rocky & ( 0 , 0 ) \\
305.01 & 6063220 & 1.79 & ( 1.75 , 1.86 ) & 4.60 & 0.76 & 0.07 & ( 0.0 , 0.2 ) \\
306.01 & 6071903 & 2.45 & ( 2.36 , 2.75 ) & 24.31 & 0.87 & 1.4 & ( 0.8 , 2.6 ) \\
307.01 & 6289257 & 1.82 & ( 1.56 , 2.01 ) & 19.67 & 1.06 & 0.06 & ( 0.0 , 0.3 ) \\
307.02 & 6289257 & 1.19 & ( 1.11 , 1.33 ) & 5.21 & 1.04 & Rocky & ( 0 , 0 ) \\
312.01 & 7050989 & 2.11 & ( 1.91 , 2.41 ) & 11.58 & 1.16 & 0.1 & ( 0.1 , 0.8 ) \\
312.02 & 7050989 & 2.15 & ( 1.91 , 2.38 ) & 16.40 & 1.18 & 0.1 & ( 0.1 , 0.8 ) \\
313.01 & 7419318 & 2.28 & ( 2.19 , 2.50 ) & 18.74 & 0.86 & 0.5 & ( 0.4 , 1.6 ) \\
313.02 & 7419318 & 1.92 & ( 1.80 , 2.11 ) & 8.44 & 0.86 & 0.1 & ( 0.1 , 0.4 ) \\
314.01 & 7603200 & 1.59 & ( 1.50 , 1.72 ) & 13.78 & 0.51 & 0.1 & ( 0.1 , 0.3 ) \\
314.02 & 7603200 & 1.43 & ( 1.35 , 1.54 ) & 23.09 & 0.52 & 0.08 & ( 0.0 , 0.2 ) \\
321.01 & 8753657 & 1.42 & ( 1.33 , 1.62 ) & 2.43 & 1.03 & Rocky & ( 0 , 0 ) \\
323.01 & 9139084 & 2.27 & ( 2.20 , 2.50 ) & 5.84 & 0.89 & 0.5 & ( 0.3 , 1.1 ) \\
327.01 & 9881662 & 1.56 & ( 1.48 , 1.71 ) & 3.25 & 1.11 & Rocky & ( 0 , 0 ) \\
333.01 & 10337258 & 2.26 & ( 2.12 , 2.91 ) & 13.29 & 1.14 & 0.3 & ( 0.2 , 2.6 ) \\
352.02 & 11521793 & 2.15 & ( 1.79 , 2.57 ) & 16.01 & 0.99 & 0.2 & ( 0.1 , 1.4 ) \\
354.01 & 11568987 & 2.59 & ( 2.49 , 2.86 ) & 15.96 & 1.04 & 1.6 & ( 1.1 , 2.7 ) \\
354.02 & 11568987 & 1.25 & ( 1.19 , 1.39 ) & 7.38 & 1.03 & Rocky & ( 0 , 0 ) \\
355.01 & 11621223 & 2.30 & ( 2.19 , 2.60 ) & 4.90 & 1.13 & 0.3 & ( 0.2 , 1.1 ) \\
361.01 & 12404954 & 1.55 & ( 1.48 , 1.76 ) & 3.25 & 0.98 & Rocky & ( 0 , 0 ) \\
369.01 & 7175184 & 1.32 & ( 1.18 , 1.71 ) & 5.89 & 1.16 & Rocky & ( 0 , 0 ) \\
369.02 & 7175184 & 1.33 & ( 1.15 , 1.67 ) & 10.10 & 1.14 & Rocky & ( 0 , 0 ) \\
385.01 & 3446746 & 2.16 & ( 1.91 , 2.59 ) & 13.15 & 1.00 & 0.1 & ( 0.1 , 1.5 ) \\
409.01 & 5444548 & 2.61 & ( 2.46 , 3.25 ) & 13.25 & 1.04 & 1.5 & ( 1.0 , 4.5 ) \\
568.01 & 7595157 & 1.58 & ( 1.44 , 2.11 ) & 3.38 & 0.89 & Rocky & ( 0.0 , 0.2 ) \\
568.02 & 7595157 & 1.05 & ( 0.96 , 1.38 ) & 2.36 & 0.87 & Rocky & ( 0 , 0 ) \\
623.01 & 12068975 & 1.36 & ( 1.31 , 1.41 ) & 10.35 & 1.11 & Rocky & ( 0 , 0 ) \\
623.02 & 12068975 & 1.33 & ( 1.29 , 1.39 ) & 15.68 & 1.11 & Rocky & ( 0 , 0 ) \\
623.03 & 12068975 & 1.16 & ( 1.14 , 1.23 ) & 5.60 & 1.11 & Rocky & ( 0 , 0 ) \\
627.01 & 4563268 & 2.43 & ( 2.34 , 2.86 ) & 7.75 & 1.17 & 0.8 & ( 0.5 , 2.3 ) \\
627.02 & 4563268 & 1.42 & ( 1.31 , 1.64 ) & 4.17 & 1.16 & Rocky & ( 0 , 0 ) \\
628.01 & 4644604 & 2.22 & ( 2.05 , 2.60 ) & 14.49 & 0.97 & 0.3 & ( 0.2 , 1.6 ) \\
632.01 & 4827723 & 1.53 & ( 1.49 , 2.01 ) & 7.24 & 0.89 & Rocky & ( 0.0 , 0.2 ) \\
639.01 & 5120087 & 2.38 & ( 2.20 , 2.94 ) & 17.98 & 1.14 & 0.5 & ( 0.4 , 2.9 ) \\
647.01 & 5531694 & 1.68 & ( 1.46 , 2.19 ) & 5.17 & 1.11 & Rocky & ( 0.0 , 0.2 ) \\
650.01 & 5786676 & 2.53 & ( 2.38 , 2.93 ) & 11.96 & 0.81 & 1.4 & ( 0.8 , 3.1 ) \\
662.01 & 6365156 & 2.25 & ( 2.10 , 2.50 ) & 10.21 & 1.16 & 0.2 & ( 0.2 , 1.1 ) \\
664.01 & 6442340 & 2.04 & ( 1.55 , 2.25 ) & 13.14 & 1.05 & 0.1 & ( 0.0 , 0.6 ) \\
664.02 & 6442340 & 1.20 & ( 1.11 , 1.45 ) & 7.78 & 1.04 & Rocky & ( 0 , 0 ) \\
664.03 & 6442340 & 1.09 & ( 1.01 , 1.30 ) & 23.44 & 1.04 & Rocky & ( 0 , 0 ) \\
665.01 & 6685609 & 2.29 & ( 2.15 , 2.98 ) & 5.87 & 1.10 & 0.3 & ( 0.2 , 2.6 ) \\
665.02 & 6685609 & 1.15 & ( 1.05 , 1.75 ) & 1.61 & 1.11 & Rocky & ( 0 , 0 ) \\
665.03 & 6685609 & 1.15 & ( 1.02 , 1.68 ) & 3.07 & 1.11 & Rocky & ( 0 , 0 ) \\
666.01 & 6707835 & 2.84 & ( 2.73 , 3.10 ) & 22.25 & 1.04 & 2.9 & ( 2.1 , 4.1 ) \\
673.01 & 7124613 & 1.76 & ( 1.72 , 2.61 ) & 4.42 & 1.14 & 0.2 & ( 0.0 , 1.1 ) \\
691.02 & 8480285 & 1.24 & ( 1.12 , 1.44 ) & 16.23 & 1.02 & Rocky & ( 0 , 0 ) \\
692.01 & 8557374 & 1.43 & ( 1.29 , 1.78 ) & 2.46 & 0.98 & Rocky & ( 0 , 0 ) \\
692.02 & 8557374 & 1.65 & ( 1.63 , 2.43 ) & 4.82 & 0.97 & 0.1 & ( 0.0 , 0.8 ) \\
693.02 & 8738735 & 2.27 & ( 2.09 , 2.80 ) & 15.66 & 1.10 & 0.3 & ( 0.2 , 2.3 ) \\
694.01 & 8802165 & 2.87 & ( 2.68 , 3.54 ) & 17.42 & 0.94 & 2.5 & ( 2.0 , 6.5 ) \\
700.02 & 8962094 & 1.45 & ( 1.41 , 2.14 ) & 9.36 & 0.91 & Rocky & ( 0.0 , 0.3 ) \\
700.03 & 8962094 & 1.39 & ( 1.30 , 1.94 ) & 14.67 & 0.93 & Rocky & ( 0.0 , 0.2 ) \\
701.01 & 9002278 & 2.26 & ( 2.18 , 2.38 ) & 18.16 & 0.66 & 0.4 & ( 0.3 , 1.4 ) \\
701.02 & 9002278 & 1.47 & ( 1.44 , 1.56 ) & 5.71 & 0.66 & Rocky & ( 0 , 0 ) \\
704.01 & 9266431 & 2.35 & ( 2.20 , 2.98 ) & 18.40 & 0.91 & 0.7 & ( 0.5 , 3.5 ) \\
708.01 & 9530945 & 2.47 & ( 2.38 , 3.29 ) & 17.41 & 1.08 & 1.1 & ( 0.8 , 4.8 ) \\
708.02 & 9530945 & 2.21 & ( 1.90 , 2.66 ) & 7.69 & 1.11 & 0.2 & ( 0.1 , 1.4 ) \\
709.01 & 9578686 & 2.30 & ( 2.15 , 2.92 ) & 21.39 & 0.89 & 0.5 & ( 0.4 , 3.1 ) \\
711.02 & 9597345 & 1.64 & ( 1.49 , 1.82 ) & 3.62 & 1.04 & Rocky & ( 0 , 0 ) \\
714.01 & 9702072 & 2.76 & ( 2.68 , 3.17 ) & 4.18 & 0.88 & 1.8 & ( 1.5 , 3.6 ) \\
717.01 & 9873254 & 2.09 & ( 1.92 , 2.40 ) & 14.71 & 1.11 & 0.2 & ( 0.1 , 1.0 ) \\
719.01 & 9950612 & 1.95 & ( 1.83 , 2.04 ) & 9.03 & 0.71 & 0.2 & ( 0.1 , 0.5 ) \\
984.01 & 1161345 & 3.05 & ( 2.97 , 3.26 ) & 4.29 & 0.91 & 3.1 & ( 2.6 , 4.1 ) \\
987.01 & 7295235 & 1.40 & ( 1.34 , 1.60 ) & 3.18 & 0.92 & Rocky & ( 0 , 0 ) \\
1002.01 & 1865042 & 1.30 & ( 1.21 , 1.58 ) & 3.48 & 1.00 & Rocky & ( 0 , 0 ) \\
1116.01 & 2849805 & 1.68 & ( 1.49 , 2.03 ) & 3.75 & 1.14 & Rocky & ( 0.0 , 0.1 ) \\
1118.01 & 2853446 & 1.58 & ( 1.49 , 2.05 ) & 7.37 & 1.02 & Rocky & ( 0.0 , 0.2 ) \\
1128.01 & 6362874 & 1.15 & ( 1.10 , 1.30 ) & 0.98 & 0.88 & Rocky & ( 0 , 0 ) \\
1150.01 & 8278371 & 1.12 & ( 1.05 , 1.51 ) & 0.68 & 1.16 & Rocky & ( 0 , 0 ) \\
1151.01 & 8280511 & 1.39 & ( 1.29 , 1.53 ) & 10.44 & 0.87 & Rocky & ( 0 , 0 ) \\
1165.01 & 10337517 & 2.29 & ( 2.10 , 2.86 ) & 7.05 & 0.94 & 0.3 & ( 0.2 , 2.2 ) \\
1216.01 & 3839488 & 1.54 & ( 1.49 , 2.17 ) & 11.13 & 1.07 & Rocky & ( 0.0 , 0.4 ) \\
1245.01 & 6693640 & 2.22 & ( 2.05 , 2.78 ) & 13.72 & 1.16 & 0.3 & ( 0.2 , 2.0 ) \\
1279.01 & 8628758 & 2.18 & ( 1.99 , 2.45 ) & 14.37 & 1.00 & 0.2 & ( 0.2 , 1.1 ) \\
1279.02 & 8628758 & 1.16 & ( 1.08 , 1.41 ) & 9.65 & 1.03 & Rocky & ( 0 , 0 ) \\
1315.01 & 10928043 & 1.55 & ( 1.45 , 1.69 ) & 6.85 & 1.14 & Rocky & ( 0 , 0 ) \\
1379.01 & 7211221 & 1.29 & ( 1.22 , 1.54 ) & 5.62 & 0.88 & Rocky & ( 0 , 0 ) \\
1438.01 & 11193263 & 1.53 & ( 1.36 , 2.24 ) & 6.91 & 1.05 & Rocky & ( 0.0 , 0.4 ) \\
1529.01 & 9821454 & 2.15 & ( 1.51 , 2.47 ) & 17.98 & 1.10 & 0.2 & ( 0.0 , 1.1 ) \\
1529.02 & 9821454 & 1.23 & ( 1.10 , 1.62 ) & 11.87 & 1.08 & Rocky & ( 0 , 0 ) \\
1531.01 & 11764462 & 1.38 & ( 1.24 , 1.87 ) & 5.70 & 1.02 & Rocky & ( 0 , 0 ) \\
1533.01 & 7808587 & 1.55 & ( 1.32 , 2.06 ) & 6.24 & 1.09 & Rocky & ( 0.0 , 0.1 ) \\
1534.01 & 4741126 & 1.42 & ( 1.42 , 2.53 ) & 20.42 & 1.20 & 0.2 & ( 0.0 , 1.3 ) \\
1534.02 & 4741126 & 1.02 & ( 0.97 , 1.62 ) & 7.64 & 1.13 & Rocky & ( 0 , 0 ) \\
1606.01 & 9886661 & 1.65 & ( 1.62 , 1.96 ) & 5.08 & 0.94 & Rocky & ( 0.0 , 0.1 ) \\
1608.01 & 10055126 & 1.55 & ( 1.48 , 2.11 ) & 9.18 & 1.05 & Rocky & ( 0.0 , 0.3 ) \\
1608.02 & 10055126 & 1.37 & ( 1.26 , 1.59 ) & 19.74 & 1.06 & Rocky & ( 0 , 0 ) \\
1628.01 & 6975129 & 2.61 & ( 2.50 , 2.94 ) & 19.75 & 1.13 & 1.8 & ( 1.1 , 3.1 ) \\
1629.01 & 8685497 & 1.43 & ( 1.31 , 1.71 ) & 4.41 & 1.15 & Rocky & ( 0 , 0 ) \\
1632.01 & 9277896 & 1.37 & ( 1.13 , 1.73 ) & 4.59 & 1.15 & Rocky & ( 0 , 0 ) \\
1738.01 & 4365645 & 1.13 & ( 1.07 , 1.48 ) & 4.17 & 0.80 & Rocky & ( 0 , 0 ) \\
1792.03 & 8552719 & 1.33 & ( 1.26 , 1.50 ) & 9.11 & 1.03 & Rocky & ( 0 , 0 ) \\
1802.01 & 11298298 & 2.43 & ( 2.35 , 3.21 ) & 5.25 & 1.09 & 0.6 & ( 0.5 , 3.6 ) \\
1806.02 & 9529744 & 1.39 & ( 1.25 , 2.18 ) & 17.93 & 1.17 & Rocky & ( 0.0 , 0.4 ) \\
1806.03 & 9529744 & 1.20 & ( 1.02 , 1.58 ) & 8.37 & 1.12 & Rocky & ( 0 , 0 ) \\
1809.01 & 8240797 & 2.32 & ( 2.12 , 2.88 ) & 13.09 & 1.17 & 0.3 & ( 0.3 , 2.6 ) \\
1809.02 & 8240797 & 1.63 & ( 1.53 , 2.43 ) & 4.92 & 1.18 & Rocky & ( 0.0 , 0.7 ) \\
1819.01 & 9597058 & 2.03 & ( 1.85 , 2.23 ) & 12.06 & 0.73 & 0.1 & ( 0.1 , 0.7 ) \\
1820.01 & 8277797 & 1.53 & ( 1.45 , 2.52 ) & 4.34 & 0.82 & Rocky & ( 0 , 1 ) \\
1837.02 & 10657406 & 1.22 & ( 1.09 , 1.70 ) & 1.68 & 0.94 & Rocky & ( 0 , 0 ) \\
1850.01 & 8826168 & 2.15 & ( 2.02 , 2.58 ) & 11.55 & 0.97 & 0.2 & ( 0.2 , 1.5 ) \\
1886.01 & 9549648 & 1.64 & ( 1.51 , 1.78 ) & 5.99 & 1.12 & Rocky & ( 0 , 0 ) \\
1893.01 & 8689793 & 1.62 & ( 1.42 , 1.96 ) & 3.56 & 0.97 & Rocky & ( 0 , 0 ) \\
1898.01 & 7668663 & 1.59 & ( 1.49 , 2.24 ) & 6.50 & 1.14 & Rocky & ( 0.0 , 0.4 ) \\
1899.01 & 7047922 & 2.26 & ( 2.07 , 2.90 ) & 19.76 & 1.14 & 0.3 & ( 0.2 , 2.7 ) \\
1909.01 & 10130039 & 1.47 & ( 1.40 , 1.85 ) & 12.76 & 1.02 & Rocky & ( 0.0 , 0.1 ) \\
1909.02 & 10130039 & 1.14 & ( 1.08 , 1.38 ) & 5.47 & 1.01 & Rocky & ( 0 , 0 ) \\
1913.01 & 9704384 & 1.44 & ( 1.38 , 1.61 ) & 5.51 & 0.95 & Rocky & ( 0 , 0 ) \\
1916.01 & 6037581 & 2.16 & ( 1.91 , 2.54 ) & 20.68 & 0.99 & 0.2 & ( 0.2 , 1.5 ) \\
1916.02 & 6037581 & 1.51 & ( 1.51 , 2.42 ) & 9.60 & 0.99 & 0.2 & ( 0.0 , 0.9 ) \\
1937.01 & 10190777 & 1.21 & ( 1.15 , 1.29 ) & 1.41 & 0.61 & Rocky & ( 0 , 0 ) \\
1955.01 & 9892816 & 2.18 & ( 1.64 , 2.60 ) & 15.17 & 1.17 & 0.2 & ( 0.0 , 1.4 ) \\
1960.01 & 6949061 & 2.19 & ( 1.56 , 2.54 ) & 8.97 & 1.13 & 0.2 & ( 0.0 , 1.1 ) \\
1960.02 & 6949061 & 2.15 & ( 1.46 , 2.46 ) & 23.22 & 1.05 & 0.1 & ( 0.0 , 1.2 ) \\
1963.01 & 10917681 & 2.23 & ( 2.08 , 2.78 ) & 12.90 & 1.02 & 0.3 & ( 0.2 , 2.2 ) \\
1972.01 & 11253711 & 2.21 & ( 2.06 , 2.80 ) & 17.79 & 1.06 & 0.3 & ( 0.2 , 2.4 ) \\
1979.01 & 7273277 & 1.00 & ( 0.96 , 1.52 ) & 2.71 & 0.75 & Rocky & ( 0 , 0 ) \\
2007.02 & 11069176 & 1.35 & ( 1.29 , 2.32 ) & 21.13 & 1.07 & 0.1 & ( 0.0 , 0.8 ) \\
2011.01 & 5384079 & 1.37 & ( 1.19 , 1.86 ) & 7.06 & 1.15 & Rocky & ( 0 , 0 ) \\
2011.02 & 5384079 & 1.10 & ( 0.99 , 1.57 ) & 17.27 & 1.21 & Rocky & ( 0 , 0 ) \\
2017.01 & 8750043 & 1.28 & ( 1.14 , 1.72 ) & 2.30 & 0.87 & Rocky & ( 0 , 0 ) \\
2026.01 & 11923284 & 1.72 & ( 1.50 , 2.03 ) & 2.76 & 1.12 & Rocky & ( 0 , 0 ) \\
2029.01 & 9489524 & 1.35 & ( 1.37 , 2.00 ) & 16.33 & 0.82 & 0.07 & ( 0.0 , 0.4 ) \\
2032.01 & 2985767 & 1.27 & ( 1.14 , 1.94 ) & 14.08 & 0.91 & Rocky & ( 0.0 , 0.2 ) \\
2033.01 & 2304320 & 1.31 & ( 1.28 , 1.71 ) & 16.54 & 0.67 & Rocky & ( 0.0 , 0.2 ) \\
2049.01 & 9649706 & 1.49 & ( 1.31 , 1.93 ) & 1.57 & 1.12 & Rocky & ( 0 , 0 ) \\
2053.01 & 2307415 & 1.55 & ( 1.56 , 2.20 ) & 13.12 & 1.09 & 0.08 & ( 0.0 , 0.4 ) \\
2053.02 & 2307415 & 1.41 & ( 1.30 , 1.64 ) & 4.61 & 1.11 & Rocky & ( 0 , 0 ) \\
2059.01 & 12301181 & 0.98 & ( 0.95 , 1.06 ) & 6.15 & 0.79 & Rocky & ( 0 , 0 ) \\
2087.01 & 6922710 & 1.37 & ( 1.30 , 1.83 ) & 23.13 & 1.05 & Rocky & ( 0.0 , 0.1 ) \\
2105.01 & 8165946 & 1.44 & ( 1.32 , 2.15 ) & 6.42 & 1.07 & Rocky & ( 0.0 , 0.2 ) \\
2110.01 & 11460462 & 1.01 & ( 0.99 , 1.73 ) & 5.04 & 1.16 & Rocky & ( 0 , 0 ) \\
2137.01 & 9364609 & 1.35 & ( 1.36 , 2.50 ) & 14.97 & 0.91 & 0.2 & ( 0.0 , 1.3 ) \\
2159.01 & 8804455 & 1.26 & ( 1.15 , 1.50 ) & 7.60 & 1.01 & Rocky & ( 0 , 0 ) \\
2246.01 & 9458343 & 1.43 & ( 1.31 , 2.23 ) & 11.90 & 1.05 & Rocky & ( 0.0 , 0.5 ) \\
2278.01 & 3342794 & 2.04 & ( 1.80 , 2.32 ) & 14.17 & 1.03 & 0.1 & ( 0.1 , 0.7 ) \\
2278.02 & 3342794 & 1.01 & ( 0.94 , 1.26 ) & 4.92 & 1.03 & Rocky & ( 0 , 0 ) \\
2281.01 & 9221517 & 0.97 & ( 0.90 , 1.23 ) & 0.77 & 0.84 & Rocky & ( 0 , 0 ) \\
2331.01 & 12401863 & 1.23 & ( 1.12 , 1.74 ) & 2.83 & 1.09 & Rocky & ( 0 , 0 ) \\
2333.01 & 11121752 & 1.18 & ( 1.10 , 1.44 ) & 3.93 & 1.07 & Rocky & ( 0 , 0 ) \\
2333.02 & 11121752 & 1.38 & ( 1.13 , 1.54 ) & 7.63 & 1.06 & Rocky & ( 0 , 0 ) \\
2342.01 & 10212441 & 1.15 & ( 1.04 , 1.37 ) & 15.04 & 1.00 & Rocky & ( 0 , 0 ) \\
2389.01 & 8494617 & 1.32 & ( 1.23 , 1.49 ) & 22.92 & 1.02 & Rocky & ( 0 , 0 ) \\
2403.01 & 2142522 & 1.20 & ( 1.10 , 1.60 ) & 13.32 & 1.07 & Rocky & ( 0 , 0 ) \\
2414.01 & 8611832 & 1.10 & ( 1.03 , 1.30 ) & 22.60 & 0.85 & Rocky & ( 0 , 0 ) \\
2443.01 & 9209624 & 1.26 & ( 1.07 , 1.61 ) & 6.79 & 1.11 & Rocky & ( 0 , 0 ) \\
2443.02 & 9209624 & 1.29 & ( 1.06 , 1.56 ) & 11.84 & 1.11 & Rocky & ( 0 , 0 ) \\
2555.01 & 5350244 & 1.24 & ( 1.05 , 1.54 ) & 12.57 & 1.17 & Rocky & ( 0 , 0 ) \\
2559.01 & 6605493 & 1.31 & ( 1.21 , 1.52 ) & 9.31 & 1.10 & Rocky & ( 0 , 0 ) \\
2563.01 & 5175024 & 1.37 & ( 1.28 , 2.27 ) & 23.48 & 1.16 & Rocky & ( 0.0 , 0.6 ) \\
2675.01 & 5794570 & 2.24 & ( 1.91 , 2.58 ) & 5.45 & 0.85 & 0.2 & ( 0.1 , 1.3 ) \\
2693.03 & 5185897 & 0.98 & ( 0.92 , 1.07 ) & 6.83 & 0.68 & Rocky & ( 0 , 0 ) \\
2711.01 & 5272233 & 1.55 & ( 1.51 , 2.48 ) & 9.02 & 1.13 & 0.2 & ( 0.0 , 0.9 ) \\
2711.02 & 5272233 & 1.42 & ( 1.36 , 2.38 ) & 17.34 & 1.09 & 0.1 & ( 0.0 , 0.8 ) \\
2730.01 & 8415200 & 1.18 & ( 1.05 , 1.61 ) & 4.52 & 1.06 & Rocky & ( 0 , 0 ) \\
2732.01 & 9886361 & 1.13 & ( 1.08 , 1.35 ) & 7.03 & 1.15 & Rocky & ( 0 , 0 ) \\
2732.02 & 9886361 & 1.25 & ( 1.16 , 1.43 ) & 13.61 & 1.15 & Rocky & ( 0 , 0 ) \\
2743.01 & 8095441 & 1.26 & ( 1.17 , 1.88 ) & 11.88 & 0.82 & Rocky & ( 0.0 , 0.1 ) \\
2906.01 & 6716545 & 1.19 & ( 1.04 , 1.51 ) & 13.91 & 1.13 & Rocky & ( 0 , 0 ) \\
2971.01 & 4770174 & 0.96 & ( 0.93 , 1.48 ) & 6.10 & 1.14 & Rocky & ( 0 , 0 ) \\
2984.01 & 7918652 & 0.98 & ( 0.95 , 1.42 ) & 11.46 & 1.13 & Rocky & ( 0 , 0 ) \\
3020.01 & 8241079 & 1.22 & ( 1.03 , 1.57 ) & 10.92 & 1.13 & Rocky & ( 0 , 0 ) \\
3075.01 & 3328080 & 0.93 & ( 0.93 , 1.49 ) & 4.77 & 0.98 & Rocky & ( 0 , 0 ) \\
3209.01 & 7017274 & 1.34 & ( 1.18 , 1.71 ) & 11.91 & 1.12 & Rocky & ( 0 , 0 ) \\
3301.01 & 8301878 & 1.26 & ( 1.14 , 2.05 ) & 20.71 & 0.97 & Rocky & ( 0.0 , 0.3 ) \\
3346.01 & 11241912 & 1.26 & ( 1.15 , 1.44 ) & 14.43 & 1.05 & Rocky & ( 0 , 0 ) \\
3384.01 & 8644365 & 1.22 & ( 1.12 , 1.43 ) & 10.55 & 1.12 & Rocky & ( 0 , 0 ) \\
3384.02 & 8644365 & 1.40 & ( 1.31 , 1.64 ) & 19.92 & 1.11 & Rocky & ( 0 , 0 ) \\
3438.01 & 6599975 & 1.27 & ( 1.12 , 2.08 ) & 14.56 & 1.17 & Rocky & ( 0.0 , 0.2 ) \\
3876.01 & 3440118 & 2.31 & ( 2.16 , 2.89 ) & 19.58 & 1.16 & 0.4 & ( 0.4 , 2.8 ) \\
3880.01 & 4147444 & 1.15 & ( 1.00 , 1.68 ) & 1.80 & 1.12 & Rocky & ( 0 , 0 ) \\
4022.01 & 7733731 & 1.06 & ( 0.95 , 1.44 ) & 4.86 & 1.06 & Rocky & ( 0 , 0 ) \\
4053.01 & 1718958 & 0.98 & ( 0.95 , 1.58 ) & 1.42 & 1.10 & Rocky & ( 0 , 0 ) \\
4320.01 & 5095082 & 0.97 & ( 0.91 , 1.28 ) & 20.66 & 0.86 & Rocky & ( 0 , 0 ) \\
4335.01 & 10730070 & 1.27 & ( 1.10 , 1.70 ) & 7.62 & 1.08 & Rocky & ( 0 , 0 ) \\
4505.01 & 8493354 & 1.27 & ( 1.04 , 1.94 ) & 18.01 & 1.20 & Rocky & ( 0.0 , 0.1 )
\enddata
\end{deluxetable*}
\clearpage

\end{document}